\documentclass{article}

\usepackage{arxiv}

\usepackage[utf8]{inputenc} % allow utf-8 input
\usepackage[T1]{fontenc}    % use 8-bit T1 fonts
\usepackage{url}            % simple URL typesetting
\usepackage{booktabs}       % professional-quality tables
\usepackage{amsfonts}       % blackboard math symbols
\usepackage{nicefrac}       % compact symbols for 1/2, etc.
\usepackage{microtype}      % microtypography
\usepackage{lipsum}
\usepackage{url}            % simple URL typesetting
\usepackage{booktabs}       % professional-quality tables
\usepackage{amsfonts}       % blackboard math symbols
\usepackage{nicefrac}       % compact symbols for 1/2, etc.
\usepackage{microtype}      % microtypography
\usepackage{graphicx}
\usepackage{booktabs} % for professional tables
\usepackage{graphicx}
\usepackage[american]{babel}
\usepackage{times}
\usepackage{helvet}
\usepackage{courier}
\usepackage{amssymb}
\usepackage{float}
\usepackage{subfigure}
\usepackage{ifthen}
\usepackage{amssymb}
\usepackage{supertabular}
\usepackage{array}
\usepackage{multirow}
\usepackage{psfrag}
\usepackage{amsmath}
\usepackage{hhline}
\usepackage{type1cm}
\usepackage{lettrine}
\usepackage{fancybox}
\usepackage{ctable}
\usepackage{bm}
\usepackage{algorithm}
\usepackage{algorithmic}
\usepackage{placeins}
\usepackage{setspace}
\usepackage{enumerate}
\usepackage{multirow}
\usepackage{enumitem}
\usepackage{url}
\usepackage{epstopdf}
\usepackage{epsfig}
\usepackage{thmtools}
\usepackage{thm-restate}
\usepackage{mathtools}
\usepackage{soul}
\usepackage{lineno}

\newcommand{\R}{\mathbb{R}}
\newcommand{\dS}[1]{d(#1, S)}
\newcommand{\dSi}[1]{d(#1, S_i)}

\newtheorem{Assumption}{Assumption}

\newenvironment{Proof}{\paragraph{Proof:}}{\hfill$\square$}

\title{distributed optimization for over-parameterized learning}

\author{
  Zhang Chi \\
  Institute of High Performance Computing\\
  Agency for Science, Technology and Research, Singapore \\
  \texttt{zhang\_chi@ihpc.a-star.edu.sg} \\
   \And
 Li Qianxiao\\
  Department of Mathematics\\
  National University of Singapore, Singapore\\
  \texttt{matlq@nus.edu.sg} \\}
\begin{document}
\maketitle

\begin{abstract}
Distributed optimization often consists of two updating phases: local optimization and inter-node communication. Conventional approaches require working nodes to communicate with the server every one or few iterations to guarantee convergence. In this paper, we establish a completely different conclusion that each node can perform an arbitrary number of local optimization steps before communication. Moreover, we show that the more local updating can reduce the overall communication, even for an infinity number of steps where each node is free to update its local model to near-optimality before exchanging information. The extra assumption we make is that the optimal sets of local loss functions have a non-empty intersection, which is inspired by the over-paramterization phenomenon in large-scale optimization and deep learning. Our theoretical findings are confirmed by both distributed convex optimization and deep learning experiments.
\end{abstract}

% keywords can be removed
%\keywords{Distributed Optimization \and Communication Efficient \and Over-Parameterization}

\section{Introduction}
\label{sec:intro}
Distributed optimization~\cite{bertsekas1989parallel} aims to optimize a global objective formed by a sum of functions:
\begin{align}
\label{eq:goal}
f(x) = \tfrac{1}{m}\sum_{i=1}^{m} f_i(x)
\qquad
f,f_i : \R^d \rightarrow \R,
\end{align}
where $m$ is the number of computational nodes and each $f_i$ is called a local loss function. Each $f_i$ may possibly be derived from the $i^\text{th}$ batch of data, and hence may differ for each node $i$.
Optimization of~(\ref{eq:goal}) usually consists of two phases: local optimization like gradient descent  and inter-node communication like model averaging. Conventional distributed optimization algorithms, such as the ``Parameter-Server'' model ~\cite{li2014scaling,ho2013more,li2014communication}, usually require immediate inter-node communication after performing local  gradient descent to simulate centralized learning and guarantee convergence for the overall model. In practice, the rapid expanding size of machine learning models, some with tens of millions of trainable parameters, often renders the communication step an increasingly significant bottleneck. This issue is further compounded by concerns on privacy and security, bandwidth requirement, power consumption and information delay~\cite{duchi2012dual,agarwal2011distributed},  where it is favorable to reduce the communication cost and only exchange information when necessary.

Among the possible solutions to alleviate the communication overhead, one  practical method is using compressed models or gradients in communication. For example, studies on quantized gradients~\cite{seide20141,alistarh2017qsgd,wangni2018gradient} and sparsified models~\cite{aji2017sparse,lin2017deep,dryden2016communication,li2018optimal} allow each node to pass low bit gradients or models to the server instead of the raw data. However, extra noise is introduced in quantization or sparsification and communication is still required at every iteration. Another possible approach~\cite{smith2018cocoa,ma2017distributed} in reducing the communication cost is to update the local models by performing $T_i > 1$ iterations of local GD before sending the model to the server. Two fundamental questions in this scenario are:

\begin{center}
	\begin{itemize}
		\item \textbf{Question 1}: \emph{Does the algorithm still converge with an arbitrary choice of $T_i$?}
		\item \textbf{Question 2}: \emph{Does more local updating $T_i$ definitely lead to less communication?}
	\end{itemize}
\end{center}

From a general distributed optimization perspective, the answers are negative. Conventional studies \cite{zhang2012communication,zhang2016parallel,zhou2017convergence,stich2018local,yu2018parallel} show that although it may not be necessary to communicate after each local GD, frequent communication after $T_i$ steps is still needed. Moreover, to ensure convergence a decaying learning rate is required or alternatively, the loss bound has a finite term that may grow with increasing $T_i$~\cite{stich2018local,yu2018parallel}. More importantly, these prior results also indicate that in general, a bigger $T_i$ leads to poorer optimization performance and may not necessarily reduce the communication cost to obtain the same overall precision. 

%In fact, their results also show that too large $T_i$ may slow down the overall convergence rate and an arbitrary large $T_i$ may even lead to divergence. To see this, consider a simple example where $d=1$, $m=2$, $f_1 = x^2$, $f_2 = (x-1)^2/2$.  With a sufficiently large $T_i$, each local model is updated towards its local optimum $x_1^*=0$ and $x_2^*=1$, and a simple averaging leads to a point around $x=1/2$ instead of the unique optimum $x^* = 1/3$ of $f$.

However, in this paper, we draw completely different conclusions from the previous results and show the answers can be positive in certain scenarios. We provide a series of theoretical analysis for convex cases and show that convergence holds for an arbitrary choice of $T_i$, and even for $T_i = \infty$ where each node updates its local model to optimality. Moreover, our results also provide answers for the second question by showing that more local updating can reduce the overall need for communication. Beyond convex cases, we show that similar conclusions may still hold by providing theoretical analysis on simple non-convex settings whose optimal sets are affine subspaces and also provide a series of experimental evidences for deep learning. These different conclusions rest upon the following intersection assumption we make throughout the paper:

\begin{Assumption}
	\label{Ass:intersection}
	Denoting $S_i := \{ x \in \R^d : f_i(x) \leq f_i(y) \ \text{holds for}\ \forall y\in\R^d \}$ as the optimal set of $f_i$, the set $S := \cap_{i=1}^m S_i$ is non-empty.
\end{Assumption}

This assumption is inspired by both the new phenomenon in modern machine learning named ``over-parameterization''~\cite{zhang2016understanding,arora2018optimization}  and
a classical mathematical problem named ``convex feasibility problem''~\cite{von1949rings}. Modern machine learning models, especially deep learning models, often consist of huge amounts of parameters that far exceed the instance numbers~\cite{ma2017power,bassily2018exponential,oymak2018overparameterized,allen2018convergence}. This over-parameterization phenomenon leads significant communication \emph{challenges} to distributed optimization as it requires enormous bandwidth to transport local models.  But in this paper,  we show that this phenomenon also brings new \emph{hopes} that allow each node to \emph{reduce the communication frequency} arbitrarily by updating its local model to (sub)optimality before sending information to server.  The underlying reason is that the training loss for over-parameterized models can often easily approach 0 due to the degeneracy of the over-parameterized functions~\cite{zhang2016understanding,du2018gradient,allen2018convergence}, indicating that there exists a common $x^*$ such that all local losses $f_i(x^*)$ are all 0 when data are distributed to multiple nodes and the above intersection assumption holds naturally.

Our work also bridges connections of the over-parameterized machine learning models with the classical convex feasibility problem~\cite{von1949rings} in mathematics, which assumes the intersection of $m$ convex closed subsets $S_i$ of a Hilbert space is non-empty and uses sequential projections~\cite{bauschke1996projection,aharoni1989block,censor2016implicit} to find a feasible point $x^* \in \cap_{i=1}^m S_i$. This non-empty intersection assumption resembles Assumption~\ref{Ass:intersection}, but  the classical approach of direct projections in convex feasible problem can often be challenging for most machine learning tasks since we cannot easily characterize local feasible sets. In this paper, we show that this projection step can be replaced by continuous local GD and convergence can still be obtained.

\textit{Notation}: \ \ Throughout this paper, we denote by $\|\cdot\|$ the Euclidean norm. If the argument is a matrix, it is the induced $2$-norm. For a closed convex set $S$, we denote by $P_S(x)$ the projection of $x$ onto $S$. The shortest distance between a point $x$ and a set $S$ is denoted by $\dS{x}$.

\section{Convex Cases}
\label{sec:convex}

To start, we first focus on the convex scenarios. Each function $f_i$ in~(\ref{eq:goal})  is assumed to be convex and $L_i$-smooth (i.e. $\nabla f_i$ is Lipschitz with constant $L_i$), and therefore the overall loss function $f$ is also convex and $L$-smooth, with $L=\tfrac{1}{m}\sum_{i=1}^{m} L_i$.  We consider the classical algorithm as seen in Alg~\ref{alg}, where each node $i$ is allowed to update $T_i$ local GD steps with step size $\eta_i$ before interacting with the server.

Denoting the point after the $n$-th communication as $x_n$, the following lemma establishes a useful bound on the evolution of the distance between $x_n$ and the common optimal set $S$.
\begin{restatable}{Lemma}{lemmaone}
	\label{lm:mother_equation}
	Consider algorithm in Alg~\ref{alg} and assume each $f_i$ is convex and $L_i$-smooth. Then,
	\begin{equation*}
	\label{eq:mother_equation}
	\dS{x_{n+1}}^2 \leq \dS{x_{n}}^2 - \tfrac{1}{m}\sum_{i=1}^{m} \sum_{t=0}^{T_i - 1} \alpha_i \| \nabla f_i (x_n^{i,t}) \|^2,
	\end{equation*}
	where 	$\alpha_i = \eta_i (\tfrac{2}{L_i} - \eta_i)$.
\end{restatable}

\begin{Proof}
	For any point $x^* \in S$,
	\begin{align}
	\label{eq:xn+1}
	\begin{split}
	\| x_{n+1} - x^* \|^2 = \| \tfrac{1}{m} \sum_{i=1}^{m} x_n^{i,T_i} - x^* \|^2
	&\leq \tfrac{1}{m} \sum_{i=1}^{m} \| x_n^{i,T_i}   - x^* \|^2
	\end{split}
	\end{align}
	For each node, we have
	\begin{align*}
	& \|  x_n^{i,t+1} - x^* \|^2 \\
	=& \|  x_n^{i,t} - x^* -  \eta_i \nabla f_i (x_n^{i,t}) \|^2  \\
	=& \|  x_n^{i,t} - x^* \|^2 - 2 \eta_i
	\langle x_n^{i,t} - x^*, \nabla f_i (x_n^{i,t})  \rangle  +(\eta_i)^2 \| \nabla f_i (x_n^{i,t}) \|^2 \\
	\leq& \|  x_n^{i,t} - x^* \|^2 - \alpha_i \| \nabla f_i (x_n^{i,t}) \|^2,
	\end{align*}
	where in the last step we used the co-coercivity of convex and $L_i$-smooth functions and the fact that $x^*$ is also in each $S_i$ due to Assumption~\ref{Ass:intersection}.
	% \begin{align*}
	% 	\langle x_n^{i,t} - x^*, \nabla f_i (x_n^{i,t}) - \nabla f_i (x^*)   \rangle \geq \tfrac{1}{L_i} \| \nabla f_i (x_n^{i,t}) - \nabla f_i (x^*)  \|^2 .
	% \end{align*}
	Summing the above from $t=0$ to $t=T_i-1$ and noticing $x_n^{i,0} = x_n$, we have
	\begin{align*}
	\|  x_n^{i,T_i} - x^* \|^2 \leq \|  x_n - x^* \|^2 - \alpha_i \sum_{t=0}^{T_i - 1} \| \nabla f_i (x_n^{i,t}) \|^2.
	\end{align*}
	Combining this with Eq.~\eqref{eq:xn+1}, we obtain
	\begin{equation*}
	\|  x_{n+1} - x^* \|^2 \leq \|  x_{n} - x^* \|^2 - \tfrac{1}{m}\sum_{i=1}^{m} \sum_{t=0}^{T_i - 1} \alpha_i \| \nabla f_i (x_n^{i,t}) \|^2.
	\end{equation*}
	Setting $x^* = P_S(x_n)$ and noticing that $\dS{x_{n+1}} \leq d(x_{n+1},x^*)$, we conclude the proof.
\end{Proof}

\begin{algorithm}[!t]
	\caption{Model Averaging for Distributed Optimization}
	\label{alg}
	\begin{algorithmic}[1]
		\item[\textbf{\underline{Worker $i=1,\cdots,m$:}}]
		\STATE{pull $x_n$ from server and initialize $x_n^{i,0} = x_n$}
		\FOR{$t=0,\cdots,T_i - 1$}
		\STATE{update $x_n^{i,t+1} = x_n^{i,t} - \eta_i \nabla f_i(x_n^{i,t}) $}
		\ENDFOR
		\STATE{push $x_n^{i,T_i}$ to server}
	\end{algorithmic}
	
	\begin{algorithmic}[1]
		\item[\textbf{\underline{Server:}}]
		\STATE{average model: $x_{n+1} = \frac{1}{m}\sum_{i=1}^m x_n^{i,T_i}$}
	\end{algorithmic}
\end{algorithm}

\paragraph{Remark:} For the general convex cases, Lemma~\ref{lm:mother_equation} already provides answers to these two questions raised in Sec~\ref{sec:intro}.

(1) To see the gradient norm converges for arbitrary $T_i$, we first notice $\{\dS{x_n}^2\}$ is a positive non-increasing sequence for $\alpha_i>0$. Therefore, for any $\delta$, there exists a sufficiently large $n$ such that
\begin{equation*}
\frac{1}{m}\sum_{i=1}^{m} \sum_{t=0}^{T_i - 1} \alpha^i \Vert \nabla f_i (x_n^{i,t}) \Vert^2 \leq \dS{x_n}^2  - \dS{x_{n+1}}^2  \leq   \delta.
\end{equation*}

The above inequality implies there exists a constant C such that $\Vert \nabla f_i (x_n^{i,t}) \Vert^2 \leq C\delta$ holds for all $i \in [1,\cdots,m]$ and $t\in [0,\cdots,T_i]$. Selecting $t=0$ and noticing $x_n^{i,0} = x_n$, we obtain
\begin{equation*}
\Vert \nabla f(x_n) \Vert^2 = \Vert\frac{1}{m}\sum_{i=1}^{m} \nabla f_i(x_n)\Vert^2 \leq \frac{1}{m}\sum_{i=1}^{m} \Vert\nabla f_i(x_n) \Vert^2 \leq C\delta.
\end{equation*}

Since $C\delta$ can be arbitrarily small, we conclude the gradient residuals $\Vert \nabla f(x_n) \Vert^2$ vanish regardless of the choice of $T_i$.

(2) To answer question 2, we notice  larger $T_i$ decreases $d(x_{n},S)$ more aggressively in Lemma~\ref{lm:mother_equation}. Namely, more local updating $T_i$ lead to less communication round $n$ to reach the same distance.

(3) The above analysis does not rely on a specific choice of learning rate $\eta_i$. Specifically, a constant learning rate $\eta_i$ can be used in local GD, and this is in contrast to studies like \cite{zhou2017convergence,stich2018local,yu2018parallel} where algorithm relies on a diminishing stepsize to obtain convergence. With a constant learning stepsize, the local learning on each node $i$ can be potentially faster, especially for  the restricted strongly convex cases considered in Sec~\ref{sec:linear_conv_sc}.

The above analysis provides qualitatively answers for the questions in Sec 1, and we shall restate these results with more concrete theorems in the following result.

\subsection{Sublinear Convergence Rate for General Convex Case}
\begin{restatable}{Theorem}{theoremtwo}
	\label{th:convergeceGuarantee}
	Suppose $f_i$ is convex and $L_i$-smooth, $\alpha_i > 0$, and $1\leq T_i \leq \infty$ for all $i=1,\dots,m$. Then,
	\begin{enumerate}
		\item[(i)] $\liminf_{n\rightarrow\infty} n \| \nabla f (x_n) \|^2 = 0$,
		\item[(ii)] $\liminf_{n\rightarrow\infty} n^{\tfrac{1}{2}}
		[
		f (x_n) - \min_{y\in\R^d} f(y)
		] = 0$.
	\end{enumerate}
\end{restatable}

\begin{Proof}
	Eq.~\eqref{eq:mother_equation} implies the sequence $\{ \dS{x_n}^2 : n\geq0 \}$ is non-increasing, and its limit exists. Define $z_n := \tfrac{1}{m}\sum_{i=1}^{m} \sum_{t=0}^{T_i - 1} \alpha_i \| \nabla f_i (x_n^{i,t}) \|^2 \geq 0$. Summing Eq.~\eqref{eq:mother_equation} and taking limit, we get
	\begin{align*}
	\lim_{n\rightarrow\infty} \sum_{k=0}^{n-1} z_k \leq \dS{x_0}^2 - \lim_{n\rightarrow\infty} \dS{x_n}^2 < \infty,
	\end{align*}
	which then implies $\liminf_{n\rightarrow\infty} n z_n \rightarrow 0$.
	Since $T_i\geq 1$, we have $$z_n \geq \tfrac{\min_i \alpha_i}{m}\sum_{i=1}^{m} \| \nabla f(x^{i,0}_n) \|^2
	\geq \min_i \alpha_i\| \nabla f(x_n) \|^2.$$ Hence $\liminf_{n\rightarrow\infty} n \| \nabla f (x_n) \|^2 \rightarrow 0.$ This proves $(i)$. To show $(ii)$, observe that for any $x^* \in S$, by convexity we have
	\begin{align*}
	f(x^*) \geq f(x_n) + \langle \nabla f(x_n), x^* - x_n \rangle,
	\end{align*}
	and so $ f(x_n) - f(x^*) \leq \| \nabla f(x_n) \| \dS{x_0}$ and $(ii)$ follows.
\end{Proof}

\paragraph{Remark:} (1) Theorem~\ref{th:convergeceGuarantee} shows that if we are working with convex functions with Lipschitz gradients, then the intersection assumption~\ref{Ass:intersection} is enough to guarantee the convergence of Alg \ref{alg} as the overall gradient norms $\| \nabla f (x_n) \|^2$ vanish, and moreover that a subsequence does so at a rate $\mathcal{O}(1/n)$; If further that the gradient norms form a monotone sequence, then this rate holds for the whole sequence. (2) Theorem~\ref{th:convergeceGuarantee} indicates that the algorithm converges for arbitrary choices of the number of local updates, including $T_i = \infty$, which represents the idealized situation where the local gradient descent problem is solved to completion before each communication step. This is attractive in practice, as it allows for each node to perform computation independently from other nodes for long times before having to exchange information.

\vspace{0.1in}
\subsection{Linear Convergence Rate for Restricted Strongly Convex Case}
\label{sec:linear_conv_sc}

The convergence rate bound in Theorem~\ref{th:convergeceGuarantee} is far from fast. In the following, we show that if some additional assumptions are adopted, then convergence can be shown to be linear in the number of communication steps.
\begin{Assumption}
	\label{Ass:restricted_sc}
	Each $f_i$ satisfies restricted strong convexity, i.e. there exists constants $\mu_i > 0$ such that
	$$ \| \nabla f_i (x) \| \geq \mu_i \dSi{x},$$
\end{Assumption}
\begin{Assumption}
	\label{Ass:separation}
	The optimum sets $\{ S_i \}$ satisfy a separation property: there exists a constant $c>1$ such that
	$$\dS{x} \leq c \cdot \tfrac{1}{m} \sum_{i=1}^{m} \dSi{x}.$$
\end{Assumption}
Assumption~\ref{Ass:restricted_sc} is a relaxed version of strong convexity, and coincides with it if $S_i$ is a singleton. Basically, this says that outside of the minimum set, the functions $f_i$ behave like strongly convex functions lower-bounded by some quadratic function. Assumption~\ref{Ass:separation} is an interesting geometric property, which roughly translates to the statement that that the ``angle of separation'' between the optimal sets is bounded from below. The fact that convergence properties depend on geometric properties of minimum sets also highlights the difference with classical distributed optimization.

\begin{restatable}{Theorem}{theoremthree}
	\label{th:restricted_sc_linear_conv}
	Let assumptions~\ref{Ass:restricted_sc} and~\ref{Ass:separation} be satisfied. Then, for any $1\leq T_i \leq \infty$ we have
	\begin{equation*}
	\dS{x_n} \leq \rho^n \dS{x_0},
	\end{equation*}
	where $\rho = \sqrt{1 - c^{-1} \min_i\{ \alpha_i\mu_i^2 \} }$ and $\alpha_i$ is such that $\alpha_i \mu_i^{2} \leq 1$.
\end{restatable}

\begin{Proof}
	From Eq.~\eqref{eq:mother_equation} and assumptions~\ref{Ass:restricted_sc} and~\ref{Ass:separation}, we have for any $T_i\geq 1$
	\begin{align*}
	\dS{x_{n+1}}^2
	\leq& \dS{x_n}^2 - \tfrac{1}{m}\sum_{i=1}^{m} \alpha_i \| \nabla f_i (x_n) \|^2 \\
	\leq& \dS{x_n}^2 - \tfrac{1}{m}\sum_{i=1}^{m} \alpha_i \mu_i^2 \dSi{x_n}^2 \\
	\leq& \dS{x_n}^2 - \min_i\{ \alpha_i\mu_i^2 \} \tfrac{1}{m} \sum_{i=1}^{m} \dSi{x_n}^2 \\
	\leq& [1 - c^{-1} \min_i\{ \alpha_i\mu_i^2 \} ] \dS{x_n}^2,
	\end{align*}
	where $c>1$. Now, observe that $\alpha_i \leq 1/L_i^2$ and $\mu_i \leq L_i$, and so $\kappa^{-1}:=\min_i(\alpha_i\mu_i^2) \in (0,1]$. In fact, $\kappa$ can be understood as an effective condition number for $f$. Thus $\rho = \sqrt{1-(c\kappa)^{-1}} \in (0,1)$ and the claim follows.
\end{Proof}

Theorem \ref{th:restricted_sc_linear_conv} shows that if each local function $f_i$ satisfies the restricted strong convexity assumption and the geometric assumption also holds, a linear convergence rate can be obtained. Similar to Theorem~\ref{th:convergeceGuarantee}, any $T_i$ including infinity guarantees convergence, which does not hold for scenarios without the intersection assumption.

\subsection{Convex Experiments}
\label{sec:convex_exp}
\subsubsection{General Convex Case}
We first validate the general convex cases where Theorem~\ref{th:convergeceGuarantee} implies the gradient residuals $\| \nabla f (x_n) \|^2$ vanish with a speed approximately of order $1/n$. The example we consider is an synthetic case from~\cite{beck2003convergence}, as the $x_n$ can approach the optimal point $x^*$ with an arbitrary slow speed. Specifically, the loss functions on two nodes are defined as $f_1(x,y) = \max^2 \left( \sqrt{x^2 + (y-1)^2}-1,0 \right)$ and $f_2(x,y)  =  \max^2 \left( y, 0 \right)$ so that the feasible set $S_1$ only intersects with $S_2$  at the point $(0,0)$. Heuristically, consider a point $x$ around the boundary of $S_1$ so that $d(x,S_1) \approx 0$. In this case, 
\begin{align*}
\frac{d(x,S)}{\frac{1}{m} \sum_{i=1}^{m} d(x,S_i)} &\approx  \frac{d(x,S)}{\frac{1}{2}  d(x,S_2)} 
= \frac{2}{\sin\theta}.
\end{align*}
If $\theta \rightarrow 0$, the left hand side goes to infinity and therefore the separation condition~\ref{Ass:separation} does not hold for this case. 

\begin{figure}[!h]
	\centering
	\includegraphics[width=0.35\linewidth]{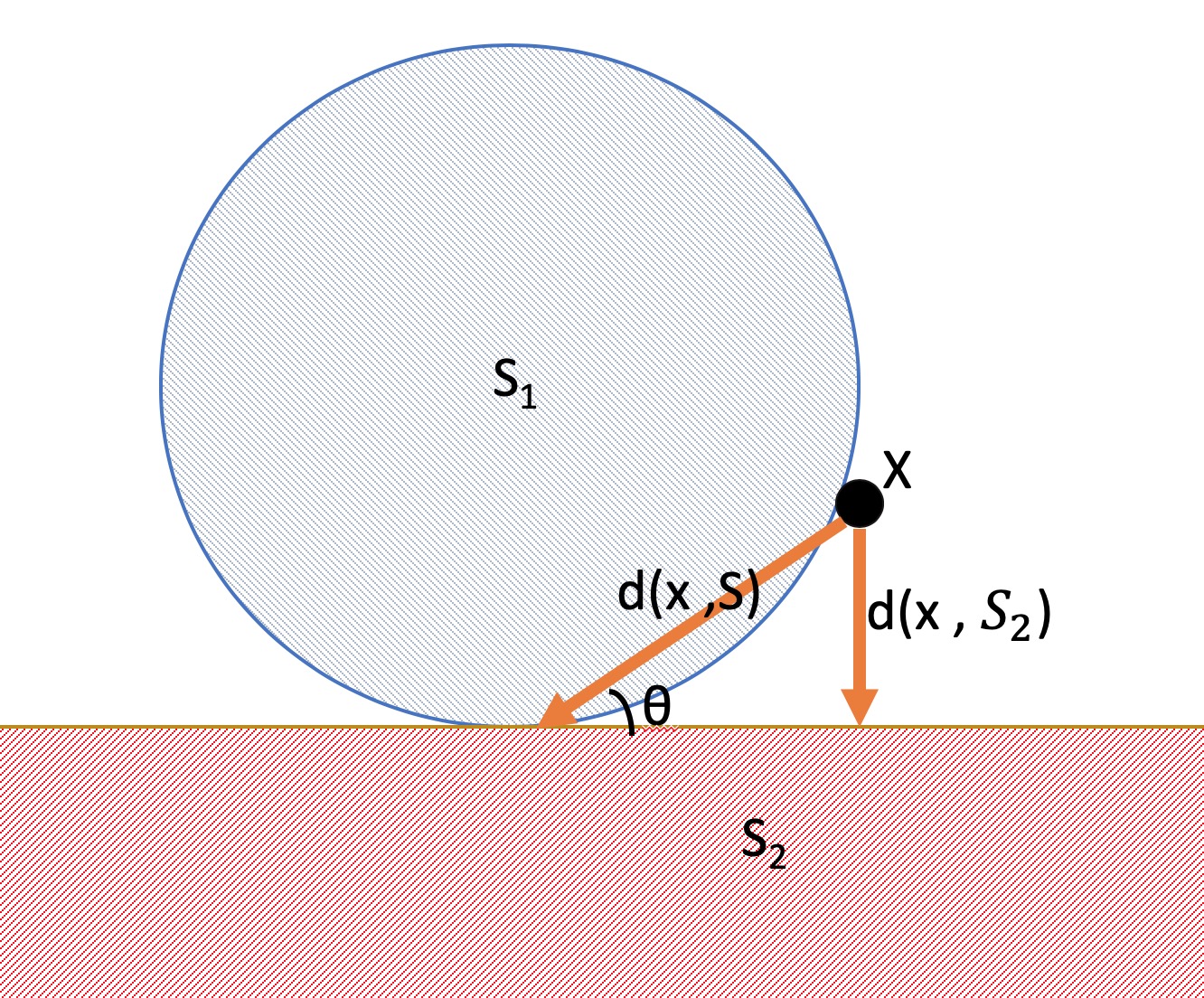}
	\caption{Synthetic experiment. The separation condition is not satisfied.}
	\label{fig:syn}
\end{figure}

%The loss functions on two nodes are defined as $f_1(x,y) = \max^2 \left( \sqrt{x^2 + (y-1)^2}-1,0 \right)$ and $f_2(x,y)  =  \max^2 \left( y, 0 \right)$ so that the feasible set $S_1$ only intersects with $S_2$  at the point $(0,0)$. 

\begin{figure*}[!t]
	\centering
	\hspace*{\fill}
	%	\subfigure[Synthetic experiment to simulate the general convex cases. The separation condition is not satisfied.]
	%	{\label{Synthetic}
	%		\includegraphics[width=.33\linewidth]{Figures/convex/crop}
	%	}
	%	\hfill
	\subfigure[Experimental results on synthetic dataset with $T_i = 10$. The black line is a reference function with $\Vert \nabla \hat{f} \Vert^2 =C/n$. Gradient residual $\| \nabla f (x_n) \|^2$ can be observed to vanish with a speed similar to $1/n$.]
	{\label{ConvexA}
		\includegraphics[width=.44\linewidth]{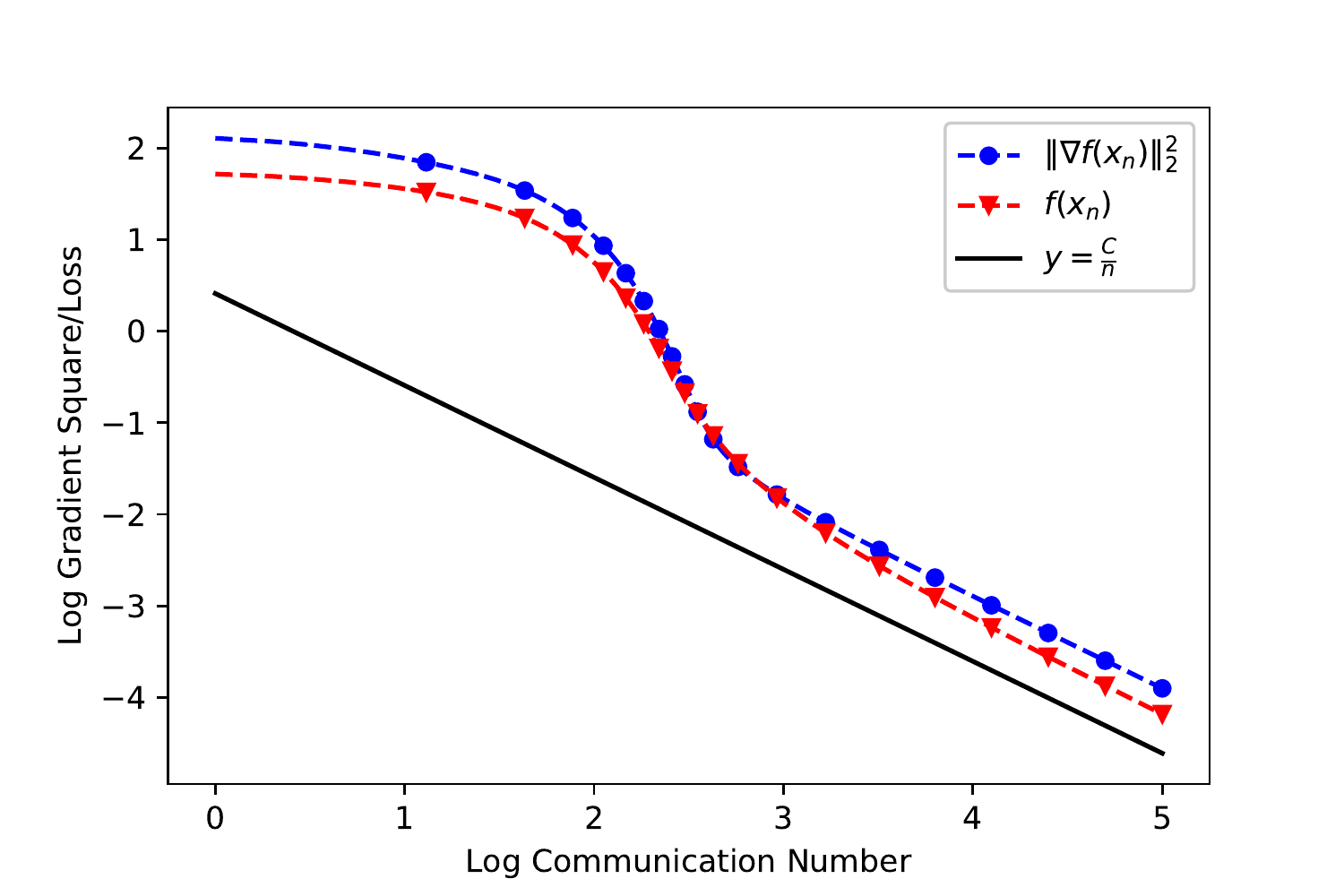}
	}
	\hfill
	\subfigure[Mean-square regression on Cancer dataset with various $T_i$. The threshold $\Vert \nabla f_i \Vert^2 \leq 10^{-8}$ is set to simulate $T_i = \infty$. Linear convergence rates can be observed for all $T_i$. ]
	{\label{ConvexB}
		\includegraphics[width=.44\linewidth]{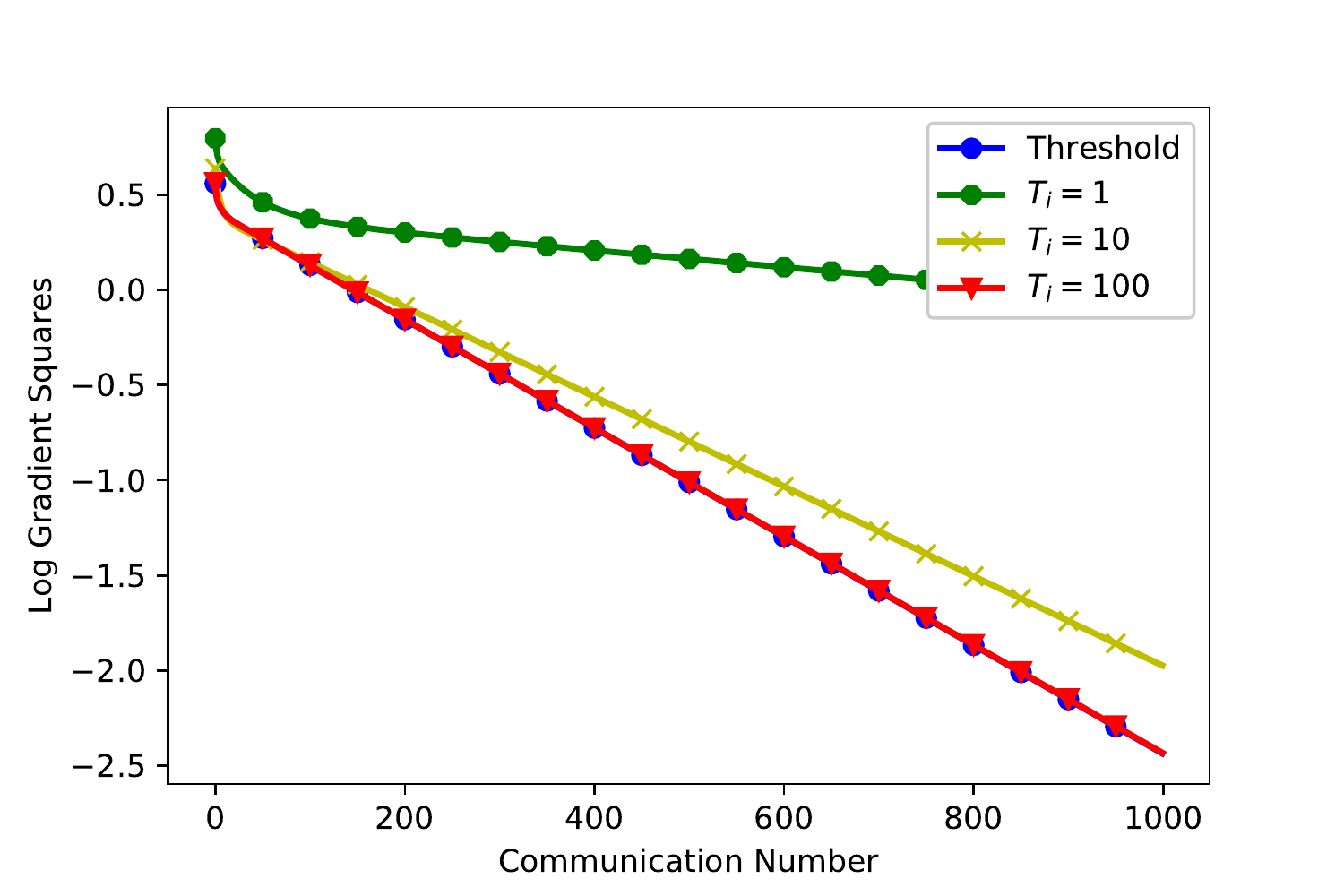}
	}
	\hspace*{\fill}
	\caption{Convex experiments. The x-axis on the left denotes $\log(n)$ and the x-axis on the right denotes $n$, and the y-axis represents the gradient residuals $\| \nabla f (x_n) \|^2$.}
	\label{fig:convex}
\end{figure*}

A start point $x_0$ is randomly selected and each node performs $T_i=10$ gradient descent steps independently before combining the parameters. Fig~\ref{ConvexA} reports how the gradient residuals $\| \nabla f (x_n) \|^2$ and the function values $f(x_n)$ vanish after each combination. To enhance visualization, we plot an auxiliary function $\hat{f}$ with a specified gradient vanishing speed $\Vert \nabla \hat{f} \Vert^2_2 =C/n$ as a reference (black line). The similar trend of the gradient residuals $\| \nabla f (x_n) \|^2$ to our reference line  validates our previous conclusion in Theorem~\ref{th:convergeceGuarantee} that  the gradient residuals $\| \nabla f (x_n) \|^2$ vanish with an approximate speed of $\mathcal{O}(1/n)$.

%, with an auxiliary function $\hat{f}$ with knowing gradient vanishing speed $\Vert \nabla \hat{f} \Vert^2_2 =C/n$ as  a reference (black line). The similar trend of the gradient residuals $\| \nabla f (x_n) \|^2$ to our reference line clearly validates our previous bounds derived for the general convex case in Theorem~\ref{th:convergeceGuarantee} that  the gradient residuals $\| \nabla f (x_n) \|^2$ vanish with a speed no worse than $\mathcal{O}(1/n)$.

\subsubsection{Linear Convergence Case}
To validate the linear convergence rate in Theorem~\ref{th:restricted_sc_linear_conv}, we perform mean-square regression on the colon-cancer dataset~\cite{alon1999broad} from LIBSVM data repository\footnote{\url{https://www.csie.ntu.edu.tw/~cjlin/libsvmtools/datasets/}}, which consists of 62 tumor and normal colon tissue samples with 2000 features. Data are evenly distributed to $m=2$ nodes, and models on each node are all over-parameterized since feature numbers far exceed instance numbers. Similar to previous experiment, each node is required to perform $T_i$ rounds of GD before communication. The infinity case $T_i = \infty$ is simulated by performing continuous GD until the local gradient norm is sufficiently small $\Vert \nabla f_i \Vert^2 \leq 10^{-8}$.

The restricted strong convex assumption and separation assumption are all satisfied in this case, and the experimental results in Fig~\ref{ConvexB} are consistent with conclusions derived in Theorem~\ref{th:restricted_sc_linear_conv}: (1) all $T_i$ leads to linear convergence rates, including the infinity case; (2) more local updating $T_i$ decreases the overall gradient residuals $\| \nabla f (x_n) \|^2$ faster, indicating sufficient local updating can reduce the overall communication cost for over-parameterized cases.

\vspace{0.2in}
\section{Beyond Convexity}
\label{sec:nonconvex}
Having established some basic convergence properties in the convex case, a natural question is whether these results hold beyond the convexity assumption. Instead of the fully general case, we first consider the following simple extension out of the convex realm, under which some convergence results can be obtained.

\subsection{Quasi-Convex Case}
Among all the non-convex scenarios, we shall first focus on the analysis of quasi-convex cases with the following assumption:
\begin{Assumption}
	\label{Ass:quasi_convex}
	Each $f_i$ is differentiable and quasi-convex, i.e. have convex sub-level-sets. Equivalently, for any $x,y\in\R^d$, $f_i(\lambda x + (1-\lambda) y ) \leq \max\{ f_i(x), f_i(y) \}$ for $\lambda \in [0,1]$.
\end{Assumption}
\begin{Assumption}
	\label{Ass:affine_opt_set}
	Each $S_i$ is an affine subspace, i.e. there exist $x^*_i \in \R^d$ and subspaces $U_i \subseteq \R^d$ such that
	\begin{align*}
	S_i = \{x^*_i\} + U_i \equiv \{ x^*_i + u: u\in U_i \}.
	\end{align*}
\end{Assumption}
Assumption~\ref{Ass:quasi_convex} is a relaxation of convexity; every convex function is quasi-convex, but the converse does not hold. For example, the sigmoid and tanh functions are quasi-convex, but not convex. Assumption~\ref{Ass:affine_opt_set} says that the optimal set of local functions are affine subspaces. This is quite a strong assumption, but it greatly simplifies the analysis. Although it is unlikely to hold in general situations, it does have some heuristic connections to neural networks, which we will subsequently discuss.

\begin{restatable}{Lemma}{lemmafour}
	\label{lm:affine}
	Let $f_i:\R^d\rightarrow \R$ satisfy assumptions~\ref{Ass:quasi_convex} and~\ref{Ass:affine_opt_set} above. Then, for every $x\in\R^d$ we have
	\begin{align*}
	f_i(x + u) = f_i(x)
	\end{align*}
	for all $u\in U_i$.
\end{restatable}
\begin{Proof}
	Without loss of generality, we may assume $\min_y f_i(y) = 0$. Let $x\in \R^d$. If $f_i(x) = 0$ then we are done. Suppose instead that $f_i(x) = c > 0$, we first show that $f_i(x+u) \leq c$ for all $u \in U_i$. Suppose for the sake of contradiction that there exists $u\in U_i$ with $f_i(x+u) = c + \delta$, $\delta>0$. Then, by continuity there exists a $\lambda \in (0,1)$ such that $f_i(\lambda(x+u) + (1-\lambda) x_i^*) = c + \delta/2$. Set $y:=\lambda(x+u) + (1-\lambda) x_i^*$ and $z := (y - \lambda x) / (1-\lambda)$, then by quasi-convexity we have
	\begin{align*}
	f_i(y) \leq \max \{ f_i(x), f_i(z) \}.
	\end{align*}
	But, by construction, $z = x^*_i + \tfrac{\lambda}{1-\lambda} u \in S_i$, and thus $f_i(z) = 0$, and so the above gives
	\begin{align*}
	c < c + \delta / 2 = f_i(y) \leq \max \{ c, 0 \} = c
	\end{align*}
	which leads to a contradiction. Hence, we must have $f_i(x+u) \leq f_i(x)$ for all $u\in U_i$, but by replacing $u$ with $-u$ we also have the reverse inequality, and so we must have $f_i(x+u) = f_i(x)$ for all $u\in U_i$.
\end{Proof}

\begin{restatable}{Lemma}{lemmafive}
	\label{corr:gd_orthogonality}
	For every $x\in \R^d$, we have $\langle u, \nabla f_i(x) \rangle = 0$ for all $u\in U_i$.
	In particular, if $\{ x^{i,t} : t\geq 0, x^{i,0} = x\}$ is a sequence of convergent gradient descent iterates under loss $f_i$, then for each $t\in [0,\infty]$, $x - x^{i,t} \perp U_i$ and $x^{i,\infty} = P_{S_i}(x)$.
\end{restatable}
\begin{Proof}
	Since $f_i$ is differentiable, we have $\langle u, f_i(x)\rangle = \lim_{\epsilon\rightarrow 0} \tfrac{1}{\epsilon} [f_i(x+\epsilon u) - f_i(x)] = 0$.
	Now, take any $u\in U_i$. We have $\langle u, x^{i,t+1} - x^{i,t} \rangle = - \eta_i \langle u, \nabla f_i(x) \rangle = 0$. Summing, we have $\langle u, (x^{i,t} - x) \rangle = 0$. If gradient descent converges, we can take the limit $t\rightarrow\infty$ to obtain $\langle u, x^{i,\infty} - x \rangle = 0$. But $x^{i,\infty} \in S_i$ which is closed since $f_i$ is continuous, and so by uniqueness of orthogonal projection, $x^{i,\infty} = P_{S_i}(x)$.
\end{Proof}

\vspace{0.2in}
Recall from Sec~\ref{sec:linear_conv_sc} that to ensure linear convergence, a geometrical assumption on the optimal sets were required. We show in the following result that in the current setting where optimal sets are affine subspaces, this separation condition is automatically satisfied.

\begin{restatable}{Lemma}{lemmasix}
	\label{lm:affine_separation}
	Let $S_i$, $i=1,\dots,m$ be a collection of affine subspaces with non-empty intersection $S = \cap_{i=1}^{m} S_i$. Then,
	\begin{align*}
	\tfrac{1}{m} \sum_{i=1}^{m} \dSi{x}
	\leq
	\dS{x}
	\leq \tfrac{c}{m} \sum_{i=1}^{m} \dSi{x}, \quad x\in \R^d,
	\end{align*}
	for some constant $c \geq 1$ with equality if and only if $S_i = S$ for all $i$.
\end{restatable}
\begin{Proof}
	The lower bound is immediate since $S\subseteq S_i$ and so $\dSi{x} \leq \dS{x}$. We now prove the upper bound. By a translation we can consider without loss of generality that $S_1,\dots,S_m$ and $S$ are subspaces.
	Thus, for each $i$ there exists a matrix $A_i$ whose rows are a orthonormal basis for $S_i^\perp$ and $\ker(A_i) = S_i$. The projection operator onto $S_i$ is then $P_i = I - A_i^\dag A_i$ (where $\dag$ denotes the Moore-Penrose pseudo-inverse) and the projection onto $S_i^\perp$ is $P_i^\perp = A_i^\dag A_i$. In particular, for each $x\in \R^d$, we have $\dSi{x} = \| P_i^\perp x \| = \| A_i^\dag A_i x \|$.
	
	Now, we show that the projection $P$ onto $S$ is $P = I - Q^\dag Q$, where $Q = I - \tfrac{1}{m} \sum_{i=1}^{m} P_i = \tfrac{1}{m} \sum_{i=1}^{m} A_i^\dag A_i$, and consequently $P^\perp = Q^\dag Q$. To show this it is enough to show that $S = \ker(\tfrac{1}{m}\sum_{i=1}^m A_i^\dag A_i)$.
	The forward inclusion is trivial, and the reverse inclusion follows from the fact that for any $x\in \ker(\tfrac{1}{m}\sum_{i=1}^m A_i^\dag A_i)$, we have
	\begin{align*}
	\begin{split}
	\| x \| = \| \tfrac{1}{m} \sum_{i=1}^{m} P_i x \|
	\leq \tfrac{1}{m} \sum_{i=1}^{m} \| P_i  x \|
	\leq \tfrac{1}{m} \sum_{i=1}^{m} \| x \| = \| x \|,
	\end{split}
	\end{align*}
	with equality if and only if $P_i x = x$ for all $i$, i.e. $x\in S$.
	
	Now, we have
	\begin{align*}
	\begin{split}
	\dS{x} =& \| P^\perp x \| = \| Q^\dag Q x \| \\
	\leq& \| Q^\dag \| \| Q x \| \\
	\leq& {\sigma_{\text{min}}(Q)}^{-1}
	\| \tfrac{1}{m} \sum_{i=1}^{m} A^\dag A x \| \\
	\leq& {\sigma_{\text{min}}(Q)}^{-1}
	\tfrac{1}{m} \sum_{i=1}^{m} \| A^\dag A x \| \\
	\leq& {\sigma_{\text{min}}(Q)}^{-1}
	\tfrac{1}{m} \sum_{i=1}^{m} \dSi{x},
	\end{split}
	\end{align*}
	where $\sigma_{\text{min}}(Q)$ denotes the smallest (non-zero) singular value of $Q$. Note that since the rows of $A_i$ are orthonormal, $\sigma_{\text{min}}(Q) \leq \tfrac{1}{m}\sum_{i=1}^{m} \| A_i^\dag A_i \| = 1$ with equality if and only if all $A_i^\dag A$ are multiples of each other (hence identical, by the orthonormal constraint), which implies $S_i=S$ for all $i$. Moreover, if $\sigma_{\text{min}}(Q) = 0$, then $Q=0$ and since each $A_i^\dag A_i$ is positive semi-definite, then $A^i = 0$ and $S_i = \R^d$ for all $i$, in which case any $c>0$ suffices. Thus, we can assume $c < \infty$.
\end{Proof}

\vspace{0.2in}
With the separation condition, we can now establish the convergence rate for the current setting.
\begin{restatable}{Theorem}{theoremseven}
	\label{th:linear_conv_affine}
	Let Assumption~\ref{Ass:quasi_convex} and \ref{Ass:affine_opt_set} hold. Take $T_i=\infty$ for all $i$ and suppose that each local gradient descent converges. Then,
	\begin{align*}
	\dS{x_n} \leq {(1 - c^{-2})}^{\frac{n}{2}} \dS{x_0}.
	\end{align*}\end{restatable}
\begin{Proof}{
		% Using Lemma, we know that $x^{i,T_i}_n = P_{S_i}(x)$
		Applying Lemma~\ref{lm:affine_separation}, we have
		\begin{align*}
		\dS{x_n}^2 \leq& \tfrac{c^2}{m} \sum_{i=1}^{m} \dSi{x_n}^2
		= \tfrac{c^2}{m} \sum_{i=1}^{m} \| x_n - P_{S_i}(x_n) \|^2.
		\end{align*}
		By Corollary~\ref{corr:gd_orthogonality}, $x^{i,\infty}_n = P_{S_i}(x_n)$, thus
		\begin{align*}
		\| x_n - x^{i,T_i}_n \|^2
		\leq& \| x_n - P_{S}(x_n) \|^2
		- \| x^{i,T_i}_n - P_{S}(x_n) \|^2
		\end{align*}
		and so,
		\begin{align*}
		\dS{x_n}^2
		\leq& c^2 \dS{x_n}^2 -
		\tfrac{c^2}{m}
		\sum_{i=1}^{m}
		\| P_{S_i}(x_n) - P_{S}(x_n) \|^2
		\\
		\leq&
		c^2 \dS{x_n}^2
		- c^2
		\|
		\tfrac{1}{m}
		\sum_{i=1}^{m}
		P_{S_i}(x_n) - P_{S}(x_n)
		\|^2 \\
		\leq&
		c^2 \dS{x_n}^2 - c^2 \dS{x_{n+1}}^2.
		\end{align*}
		Rearranging, we have
		\begin{align*}
		\dS{x_{n+1}} \leq \sqrt{1 - c^{-2}} \dS{x_{n}}.
		\end{align*}}
\end{Proof}

\paragraph{Remark:} Gradient descent for quasi-convex function can be arbitrary slow and therefore
selecting a small $T_i$ may also lead to arbitrary slow convergence rate. In the above theorem, we  set $T_i= \infty$ to make sure the local model get sufficient updates and hence the overall convergence is guaranteed.

\clearpage
\subsection{Deep Learning}
As alluded to earlier, although Assumption~\ref{Ass:affine_opt_set} is quite restrictive, it represents an interesting class of problems where gradient descent and projections are intimately connected. Moreover, deep learning models typically consist of nested affine transformations and nonlinear activations. In the over-parameterized setting where hidden node numbers are large, the affine transformations create degeneracies exactly in the form of affine subspaces.

Of course, in general the optimal sets of deep learning loss functions may be unions of affine subspaces, and furthermore the loss functions need not be quasi-convex, and hence the result above does not directly apply to deep neural networks. But we still get some \textbf{motivations} that in deep learning scenarios, the answers for these two questions in Sec~\ref{sec:intro} may be more similar to over-parameterized convex cases instead of the conventional distributed studies~\cite{li2014communication,stich2018local,yu2018parallel} that require $T_i$ to be sufficiently small to guarantee convergence. On the contrary, updating local models more precisely (with large or even infinite $T_i$) can indeed reduce the overall communication cost, as the optimal sets of these local models are likely to intersect.

\textbf{\begin{figure*}[!t]
		\centering
		\subfigure[Gradient residual $\Vert \nabla f(x_n) \Vert^2$ for the Non-Intersected case.]
		{\label{nonInterA}
			\includegraphics[width=.42\linewidth]{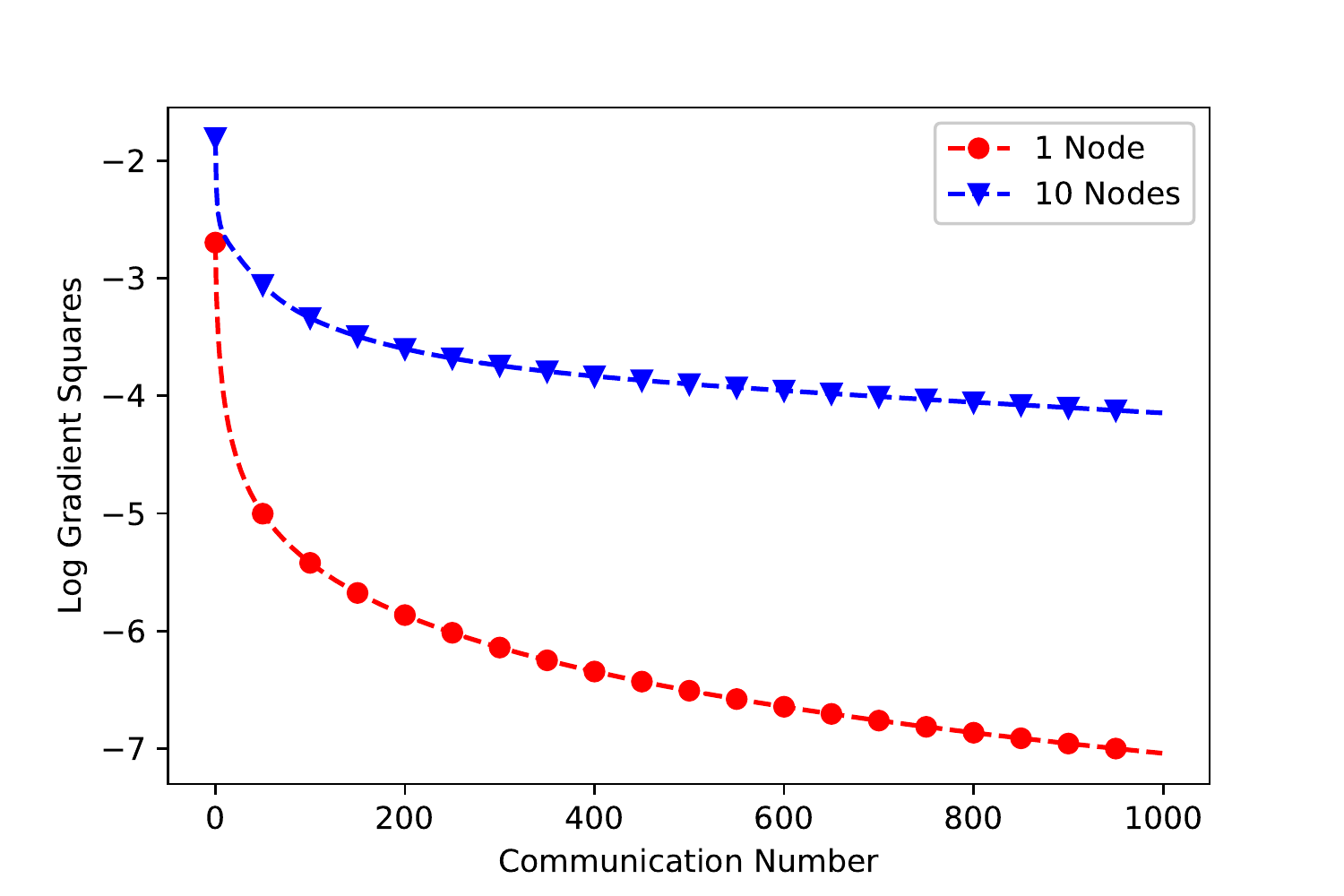}
		}
		\subfigure[Loss for the Non-Intersected case.]
		{\label{nonInterB}
			\includegraphics[width=.42\linewidth]{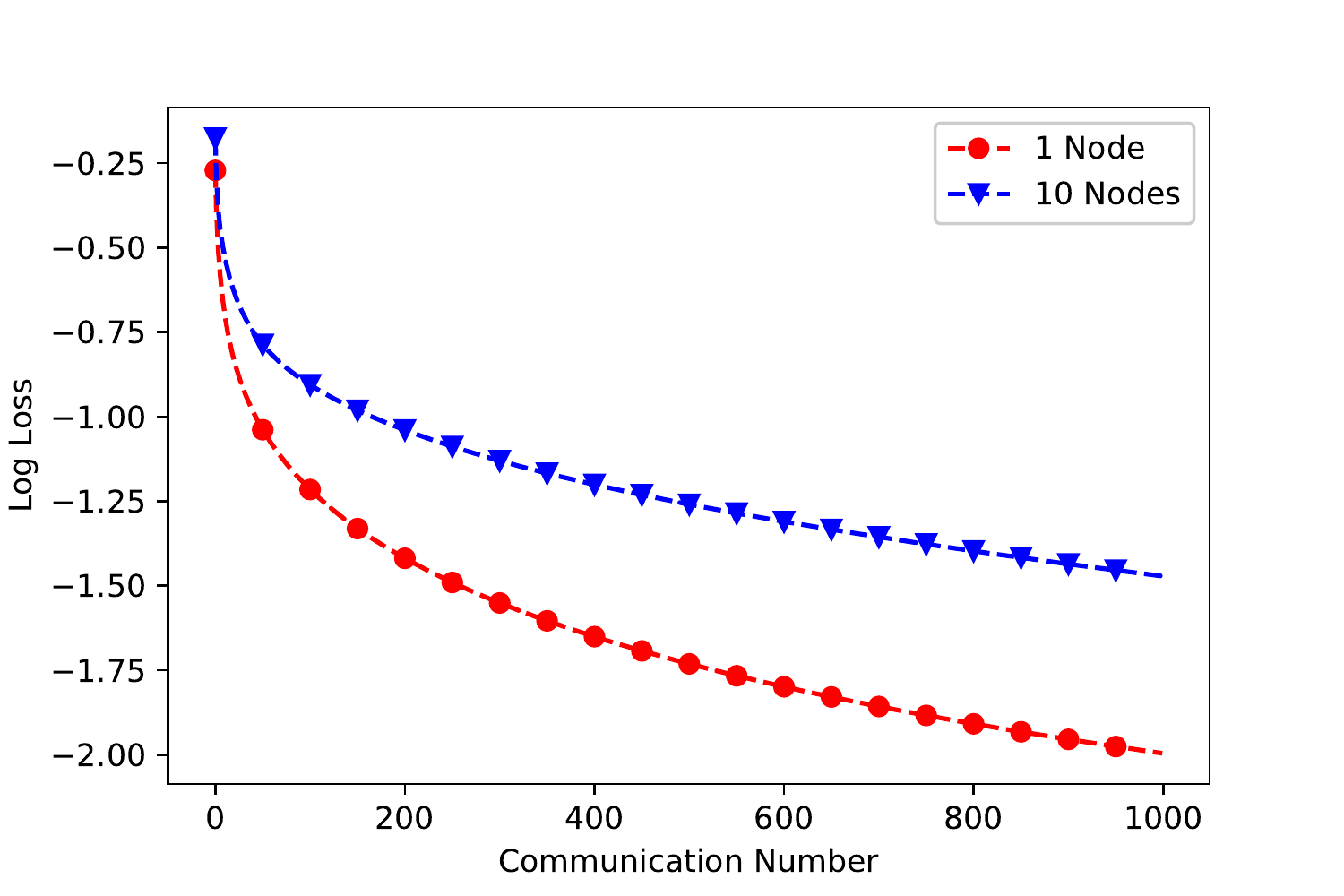}
		}
		\subfigure[Gradient residual $\Vert \nabla f(x_n) \Vert^2$ for the Intersected case.]
		{\label{InterA}
			\includegraphics[width=.42\linewidth]{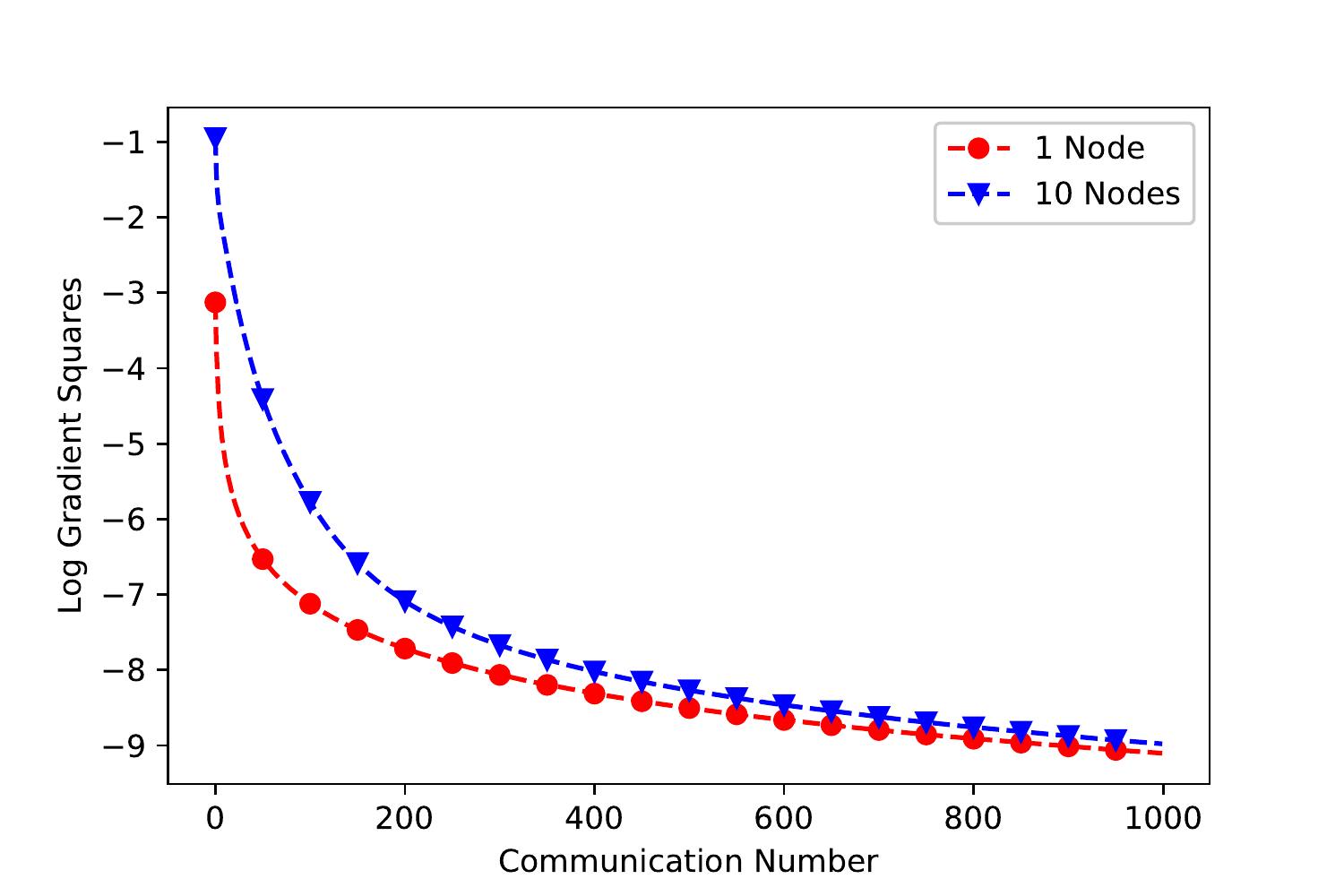}
		}
		\subfigure[Loss for the Intersected case.]
		{\label{InterB}
			\includegraphics[width=.42\linewidth]{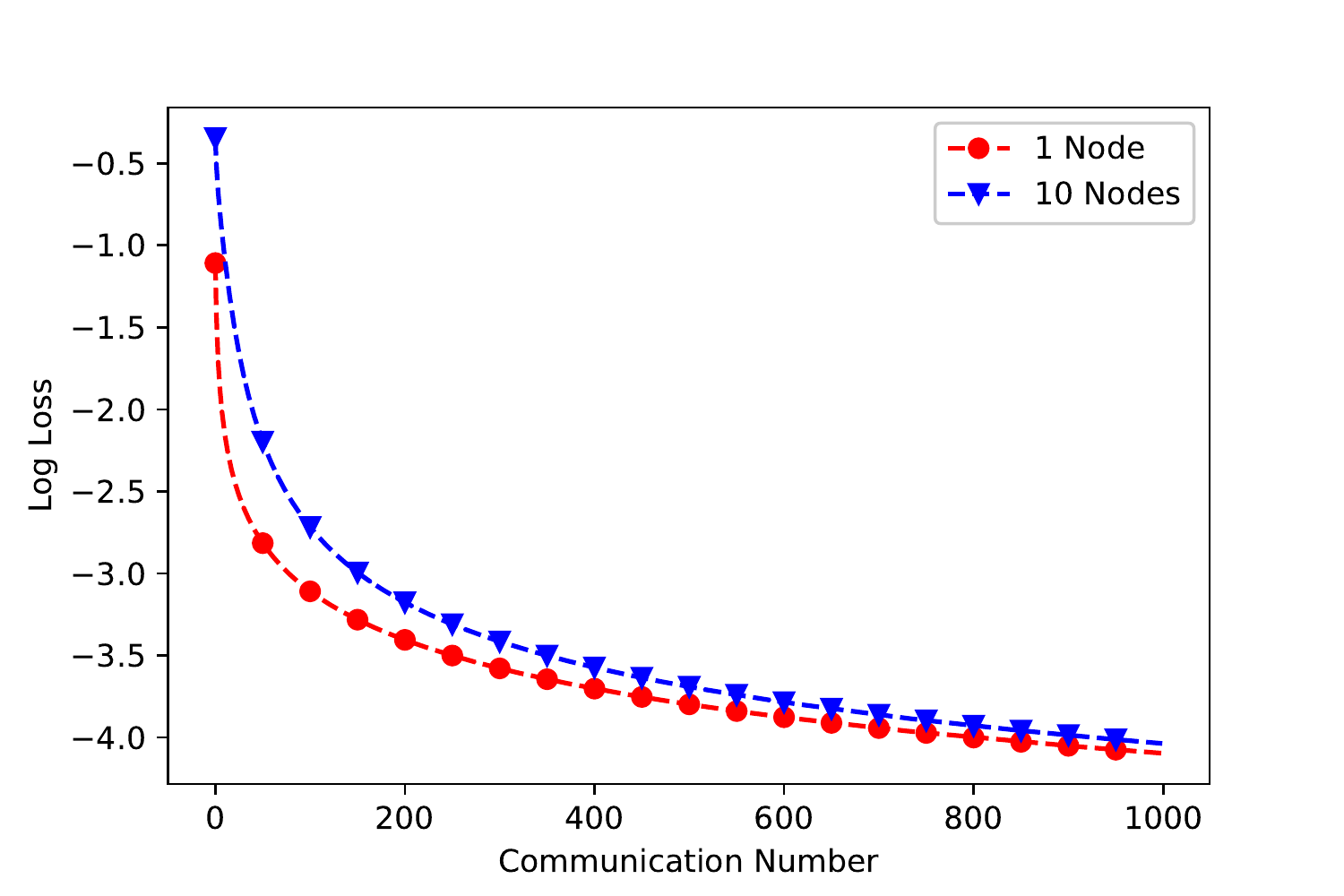}
		}
		\caption{1 Layer Neural Network on MNIST dataset. $T_i =100$ for all cases.}
		\label{fig:1LayerNN}
\end{figure*}}

\subsubsection{Necessity of the Intersection Assumption}
Before further numerical validation, we first highlight the necessity of the intersection assumption~\ref{Ass:intersection} that  distinguishes our work from previous studies. 

We select the first 500 training samples from MNIST dataset~\cite{lecun1998gradient} and construct two 1-layer neural networks:  (1) the first is to directly transform the $28 \cdot 28$ image into 10 categories by an affine transformation followed by softmax cross-entropy loss, which we name as the ``Intersected Case'' since the parameters exceed the instance numbers; (2) the second is to perform two continuous max-pooling twice with a $(2,2)$ window before the final prediction, which we name as the ``Non-Intersected Case'' since the total parameters number is 490 and the intersection assumption is not satisfied.

Figure~\ref{fig:1LayerNN} shows the results of centralized training only on the server~(denoted as ``1 Node'') and distributed training on 10 nodes. Without the intersection condition being satisfied, the gradient residuals $\Vert \nabla f(x_n) \Vert^2$ may not even vanish on the ``Non-Intersected case'' in Fig~\ref{nonInterA} and the distributed training loss $f(x_n)$ can also be different from  centralized learning in Fig~\ref{nonInterB}.   On the contrary, both the gradient residuals and the loss on 10 learning nodes perform in a similar way to centralized learning for the ``Intersected case'' in Fig~\ref{InterA} and ~\ref{InterB}. These different results validate the importance of the intersection assumption we made in the previous part.

\subsubsection{LeNet and ResNet} 
In practice, most deep learning models are highly over-parameterized, and the intersection assumption is likely to hold. In these scenarios, we numerically test the performance of Alg \ref{alg} on non-convex objectives and explore whether large or even ``infinite'' $T_i$ leads to less communication requirement. To do so, we select two classical benchmarks for deep learning: LeNet~\cite{lecun1998gradient} on MNIST dataset and ResNet-18~\cite{he2016deep} on CIFAR-10 dataset~\cite{krizhevsky2009learning}. To accelerate experiments, we only select the first 1000 training samples from these two dataset (although we also provide experiments on the complete dataset in appendix), and evenly distribute these instances to multiple learning nodes. Similar to our previous convex experiments, each node is required to perform $T_i$ iterations of GD before sending models to server, and the $T_i = \infty$ is simulated by continuous gradient descents until local gradient residual $\Vert \nabla f_i \Vert^2$ is sufficiently small.

\begin{figure*}[!t]
	\centering
	\hspace*{\fill}
	%	\subfigure[Synthetic experiment to simulate the general convex cases. The separation condition is not satisfied.]
	%	{\label{Synthetic}
	%		\includegraphics[width=.33\linewidth]{Figures/convex/crop}
	%	}
	%	\hfill
	\subfigure[LeNet for MNIST dataset. The threshold is set as $\Vert \nabla f_i \Vert_2^2 \leq 10^{-4}$.]
	{\label{LeNet}
		\includegraphics[width=.44\linewidth]{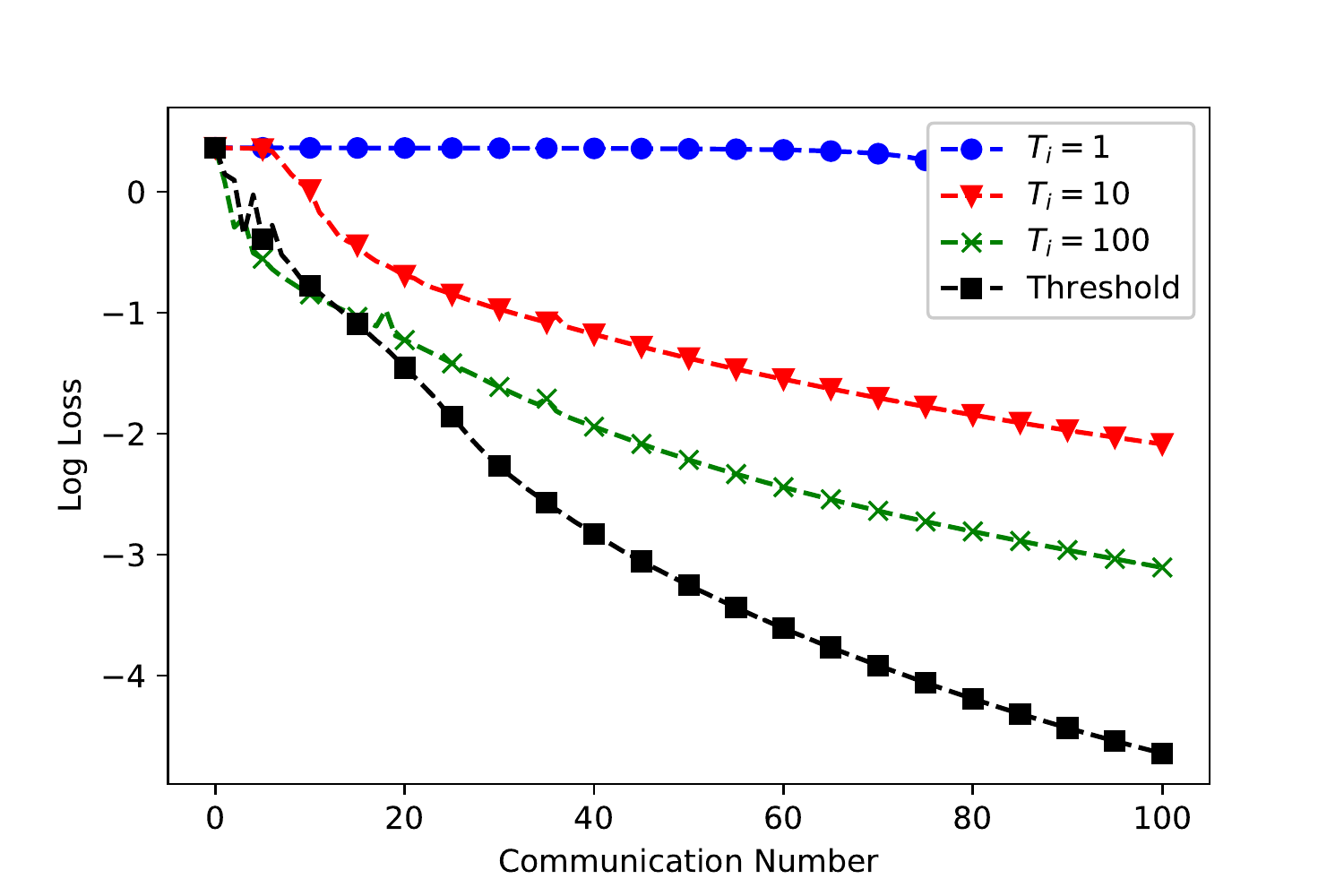}
	}
	\hfill
	\subfigure[ResNet for CIFAR 10 dataset. The threshold is set as $\Vert \nabla f_i \Vert_2^2 \leq 10^{-2}$.]
	{\label{ResNet}
		\includegraphics[width=.44\linewidth]{10Nodes_Loss.pdf}
	}
	\hspace*{\fill}
	\caption{Deep learning experiments. The x-axis  denotes the communication round $n$ and the y-axis denotes the $\log(f)$. }
	\label{fig:dl}
\end{figure*}

Figure~\ref{fig:dl} shows the experimental results on these benchmarks. The result is consistent with our previous convex experiments that choices for $T_i$ is no longer limited as the conventional studies, and larger $T_i$ decreases the total loss more aggressively. In other words, updating local model more precisely can reduces communication cost for these two deep learning models. Note that for ResNet-18, we intentionally set the local gradient norm threshold to a relatively small number $10^{-2}$, and hence the ``Threshold'' method requires thousands of epochs to reach this borderline in the beginning but only need a few epochs after 40 iterations, which explains why it first outperforms $T_i=100$ but then is inferior to it.

\section{A Quantitative Analysis of the Trade-off between Communication and Optimization}
\label{sec:tradeoff}

In previous sections we have focused on convergence properties. Recall that we proved in the convex case, essentially any frequency of local updates is sufficient for convergence. This effect then brings into relevance the following important practical question: given a degenerate distributed optimization problem, can we decide how many steps $T_i$ to take locally before a combination, in order to optimize performance or minimize some notion of computational cost? Note that this question is not well-posed unless convergence is guaranteed for any (or at least a large range of) $T_i$, as we have established in Sec 3 by relying on the degeneracy assumption.

Building on the results earlier, we now show that in our setting, this question can be answered quantitatively and this gives guidance to designing efficient distributed algorithms.
From Lemma~\ref{lm:mother_equation}, it is clear that the decrement of $\dS{x_{n}}$ come from two sources, one from the frequency of applying the outer iteration in $n$ (communication), and the other from the size of $\sum_{t=0}^{T_i - 1} \alpha_i \| \nabla f_i (x_n^{i,t}) \|^2$, which relies on $T_i$ and also the rate of convergence of local gradient descent steps (optimization). It is well-known that for general smooth convex functions, the upper-bound for the decay of gradient norms is $\mathcal{O}(t^{-1})$. However, depending on the loss function at hand, different convergence rates can occur, ranging from linear convergence to sub-linear convergence in the form of power-laws. Therefore, to make headway one has to assume some decay rate of local gradient descent.
To this end, let us assume that the local gradient descent decreases the gradient norm according to
\begin{align}
\label{eq:h_def}
\| \nabla f_i(x_n^{i,t}) \|^2 \geq h_i(t) \| \nabla f_i(x_n^{i,0}) \|^2
\end{align}
where $h_i(t)$ is a positive, monotone decreasing function with $h_i(0)=1$.

Let $\epsilon>0$ be fixed and define
\begin{align}
\label{eq:n_star_def}
n^* = \inf \{ k\geq 0, \|\nabla f(x_k)\|^2 \leq \epsilon \}.
\end{align}
From Lemma~\ref{lm:mother_equation} and Eq.~\eqref{eq:h_def} we have
\begin{align*}
\dS{x_{n+1}}^2 \leq \dS{x_{n}}^2
- \tfrac{1}{m}\sum_{i=1}^{m} \sum_{t=0}^{T_i - 1} \alpha_i h_i(t) \| \nabla f_i( x_n) \|^2.
\end{align*}
Assume for simplicity $T_i=T$ for all $i$. we have for each $n \leq n^* - 1$,
\begin{align*}
\dS{x_{n+1}}^2 \leq \dS{x_{n}}^2
- \sum_{t=0}^{T - 1} \alpha h(t) \epsilon,
\end{align*}
where $\alpha := \min_i \alpha_i$ and $h(t) := \min_i h_i(t)$.
Hence,
\begin{align}
\label{eq:n_star_expr}
n^* \leq \tfrac{\dS{x_0}^2}{ \alpha \epsilon \sum_{t=0}^{T-1} h(t)}.
\end{align}
This expression concretely links the number of steps required to reach an error tolerance to the local optimization steps.

Now, we need to define some notion of cost in order to analyze how to pick $T$. In arbitrary units, suppose each communication step has associated cost $C_c$ per node and each local gradient descent step has cost $C_g$. Then, the total cost for first achieving $\|\nabla f(x_n) \|^2 \leq \epsilon $ is
\begin{align*}
\begin{split}
C_{total} &= (C_c m + C_g m T) n^* \\
&= C_c m ( 1 + r T ) n^* \\
&\leq C_c m \dS{x_{0}}^2 {(\alpha\epsilon)}^{-1} \tfrac{( 1 + r T )}{ \sum_{t=0}^{T-1} h(t)},
\end{split}
\end{align*}
where we have defined $r := C_g / C_c$. We are mostly interested in the regime where $r$ is small, i.e. communication cost dominates gradient descent cost.

The key question we would like to answer is: for a fixed and small cost ratio $r$, how many gradient descent steps should we take for every communication and combination step in order to minimize the total cost?
It is clear that the answer to this question depend on the behavior of the sum $\sum_{t=0}^{T-1} h(t)$ as $T$ varies. Below, let us consider two representative forms of $h(t)$, which gives very different optimal solutions.

\paragraph{Linearly Convergent Case.}
We first consider the linearly convergent case where $h(t) = \beta^T$ and $\beta \in (0,1)$. This is the situation if for example, each $f_i$ is strongly convex (in the restricted sense, see Assumption~\ref{Ass:restricted_sc}). Then, we have
\begin{align*}
\sum_{t=0}^{T-1} h(t) = \tfrac{1 - \beta^T}{1 - \beta},
\end{align*}
and so
$
C_{total} \leq
C_c m \dS{x_0}^2 (1 - \beta)
{(\alpha\epsilon)}^{-1}
\tfrac{1 + r T}{1 - \beta^T}.
$

The upper bound is minimized at $T=T^*$, with
\begin{align*}
T^* =
\tfrac{1}{\log \beta}
\left[
1 + W^-(-e^{-1} \beta^{\tfrac{1}{r}})
\right]
- \tfrac{1}{r},
\end{align*}
where $W^-$ is the negative real branch of the Lambert's $W$ function, i.e. $W^{-}(x e^x) = x$ for $x \in [-1/e, 0)$. For small $x$, $W^-$ has the asymptotic form $W^- = \log (-x) + \log (- \log (-x) ) + o(1)$. Hence, for $r\ll 1$ we have
\begin{align*}
T^* =
\log
\left(
1 + \tfrac{\log (\beta^{-1})}{ r }
\right)
+ o(1).
\end{align*}

\paragraph{Sub-linearly Convergent Case.}
Let us suppose instead that $h(t) = 1 / (1+ a t)^\beta$ for some $a>0,\beta >1$. This is a case with sub-linear (power-law) convergence rate, and is often seen when the local objectives are not strongly convex, e.g. $x^{2l}$ where $l$ is a positive integer greater than 1. For instance, in this case one can show that for small learning rates, $h(t)$ is approximately of this form with $a=2l - 2$ and $\beta = (2l - 1) / (2l-2)$

By integral comparison estimates we have
\begin{align*}
\int_{0}^{T} h(s) ds \leq \sum_{t=0}^{T-1} h(t) \leq 1 + \int_{0}^{T-1} h(s) ds.
\end{align*}
Therefore,
\begin{align*}
\tfrac{1 - {(1+a T)}^{1-\beta}}{a(\beta-1)}
\leq
\sum_{t=0}^{T-1} h(t)
\leq
1 +     \tfrac{1 - {(1+a (T-1))}^{1-\beta}}{a(\beta-1)}.
\end{align*}
Thus, we have
\begin{align*}
C_{total} \leq
C_c m \dS{x_0}^2 a (\beta - 1)
{(\alpha\epsilon)}^{-1}
\tfrac{1 + r T}{1 - {(1 + a T)}^{1-\beta}}.
\end{align*}
The minimizing $T^*$ is the unique positive solution of the algebraic equation
\begin{align}
r \left((1 + a T^*)^{\beta }-1\right)-a (\beta +\beta  r T^*-1) = 0,
\end{align}
whose asymptotic form for $r\ll 1$ is
\begin{align*}
T^* = \tfrac{1}{a} \left(
\left[
\tfrac{a(\beta-1)}{r}
\right]^{\tfrac{1}{\beta}}
- 1
\right)
+
o(r^{-\tfrac{1}{\beta}}).
\end{align*}

From the explicit calculations above, we can see that the number of local gradient descent steps that minimizes the upper-bound total cost depends in a non-trivial manner on the speed of local gradient descent. If the latter is fast (linear convergence case), then the number of local steps to take is small $T^*\sim \log(1/r)$, whereas if local gradient descent is slow (sub-linear convergence case), one should take more local steps and $T^*\sim 1/r^{1/\beta}$. Besides theoretical interest, these estimates can be used in practice to tune the number of local descent steps: one may detect the order of local convergence on the fly, then use these estimates as a guideline to adjust $T$. This gives a principled way to balance optimization and communication, and is potentially useful for solving practical large-scale problems.

\paragraph{Experiment.}
\begin{figure}
	\centering
	\includegraphics[width=0.45\linewidth]{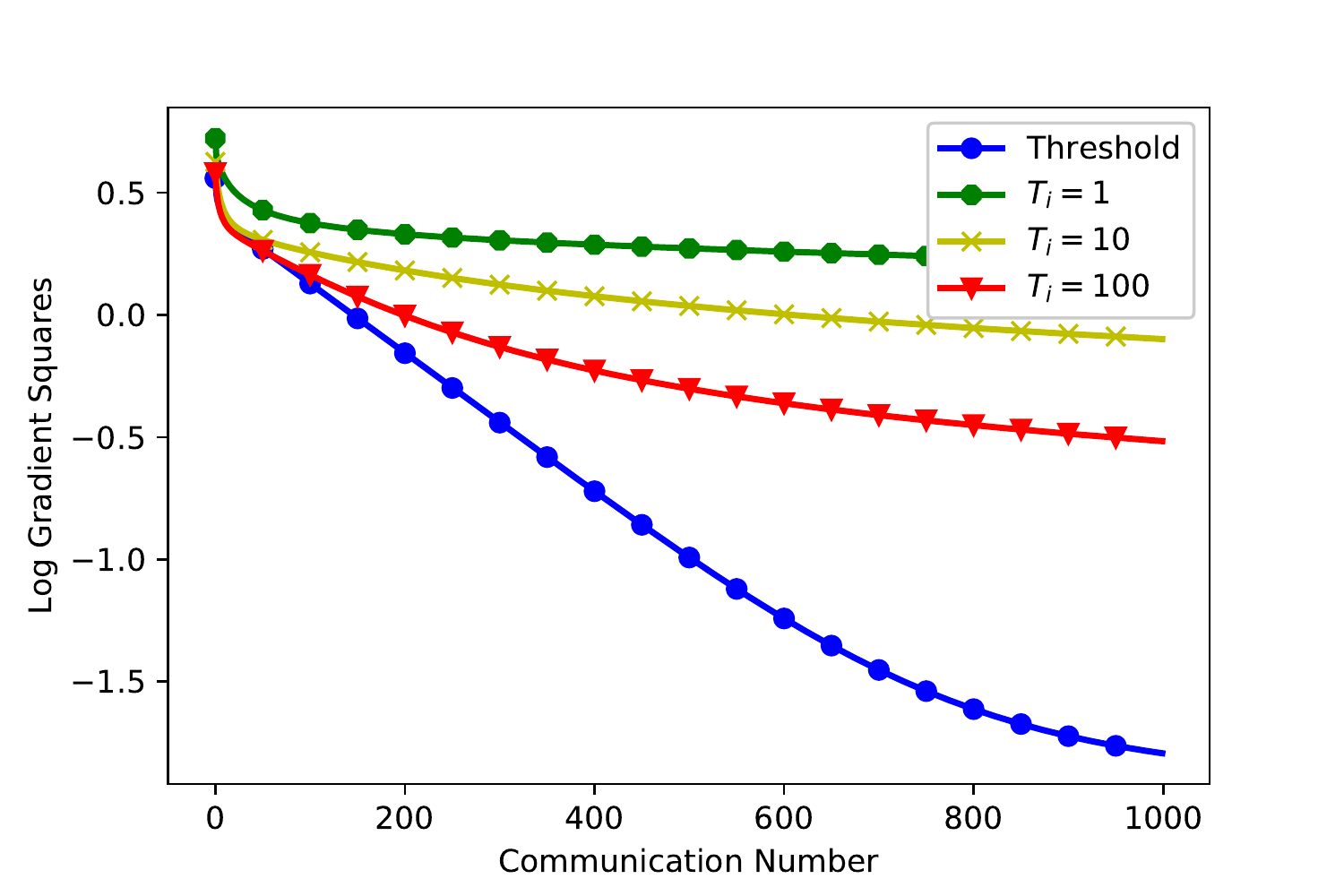}
	\caption{Cancer dataset with quartic loss function. The threshold is set as $\Vert \nabla f_i \Vert_2^2 \leq 10^{-8}$.}
	\label{fig:quartic}
\end{figure}
For comparison, we replace the quadratic loss  in Fig \ref{ConvexB} with quartic loss and obtain Fig \ref{fig:quartic}. Gradient for quadratic loss starts with a large initial value but vanishes exponentially, and $T_i=100$ already coincide with threshold case in \ref{ConvexB}. Or equivalently, each node only need to maintain a relatively small $T_i$ and then refill gradient through combination. In contrary, gradient for quartic loss begins with a smaller value and decreases as a sub-linear case. Therefore, a sufficient large $T_i$ is required to reduce total communication rounds.

\section{Conclusion}
\label{sec:conclusion}
In this paper, we analyzed the dynamics of distributed gradient descent on degenerate loss functions where the optimal sets of local functions intersect. The motivation for this assumption comes from over-parameterized learning that is becoming increasingly relevant in modern machine learning. We showed that under convexity and Lipschitz assumptions, distributed gradient descent converges for arbitrary number of local updates before combination. Moreover, we show that the convergence rate can be linear under the restricted convexity assumption, and that the convexity conditions can be relaxed if the optimal sets are affine subspaces -- an assumption connected in spirit to the degeneracies that arises in deep learning. Lastly, we analyzed quantitatively the trade-off between optimization and communication, and obtained practical guidelines for balancing the two in a principled manner.

\bibliographystyle{unsrt}  
\bibliography{DOP.bib} 

\clearpage
\appendix
\section{Supplementary Experiments}
\subsection{LeNet Experiments with Various Node Numbers}
\begin{figure*}[!h]
	\centering
	\subfigure[Loss for 2 nodes]
	{\label{}
		\includegraphics[width=.42\linewidth]{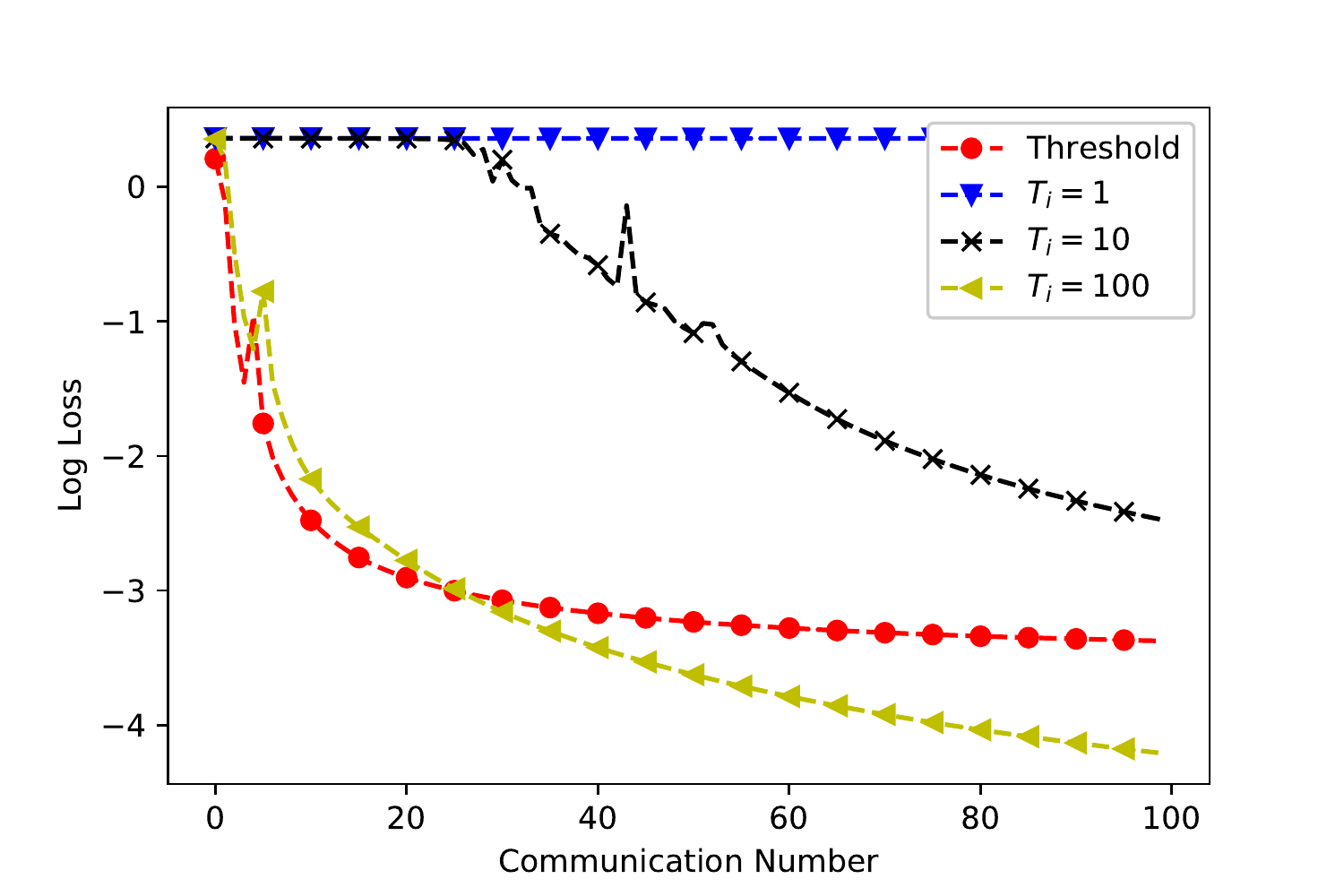}
	}
	\subfigure[Accuracy for 2 nodes]
	{\label{}
		\includegraphics[width=.42\linewidth]{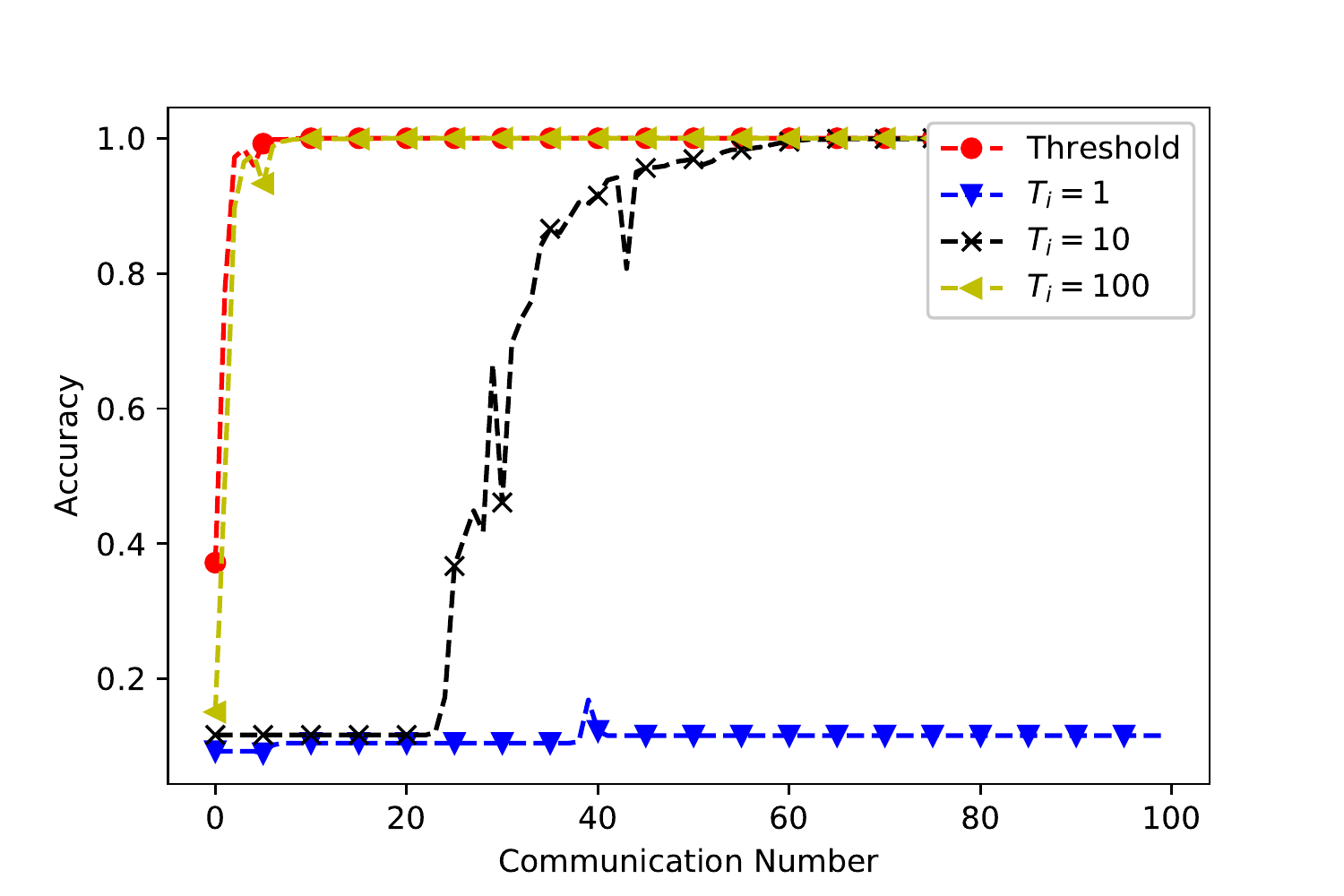}
	}
	\subfigure[Loss for 5 nodes.]
	{\label{}
		\includegraphics[width=.42\linewidth]{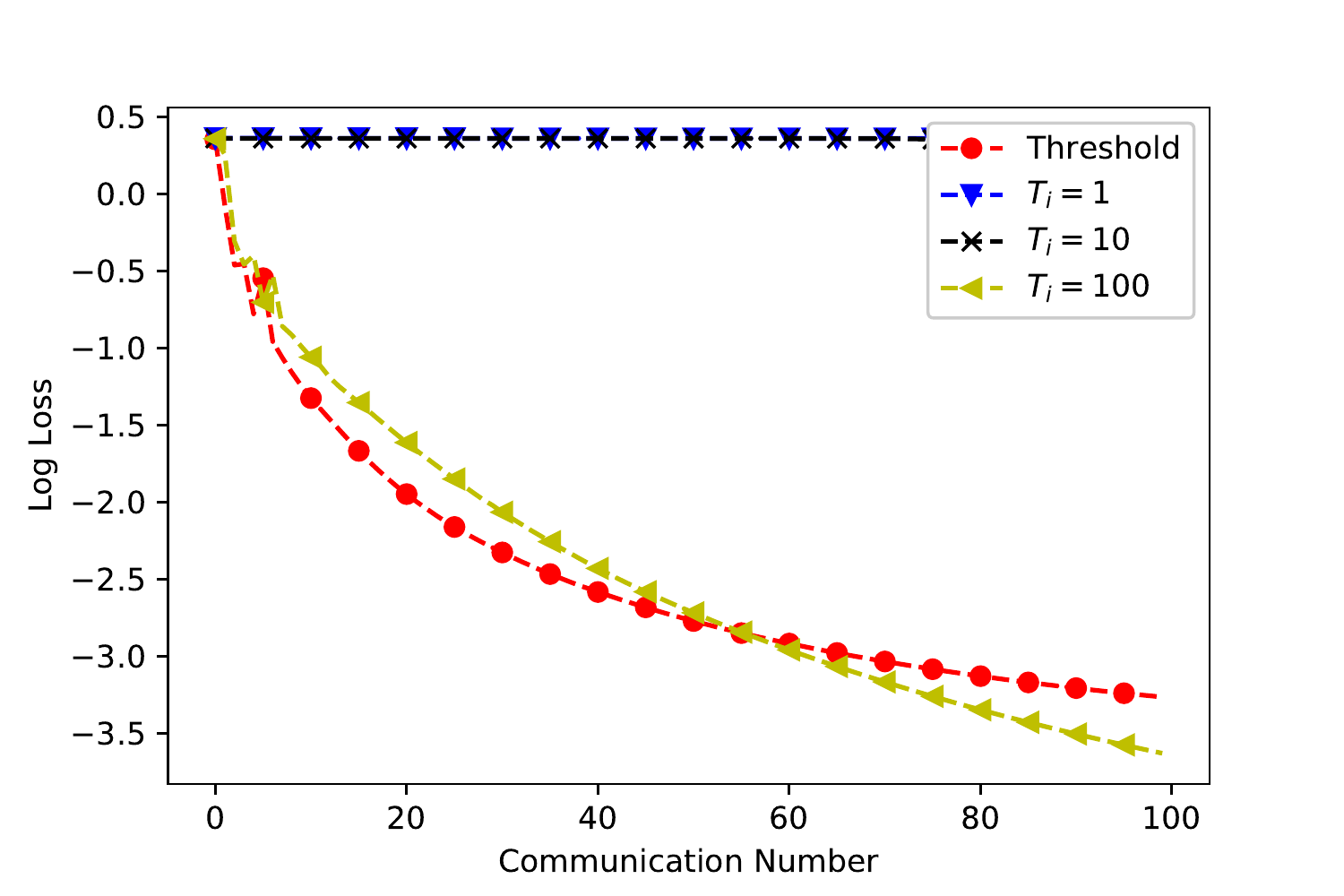}
	}
	\subfigure[Accuracy for 5 nodes.]
	{\label{}
		\includegraphics[width=.42\linewidth]{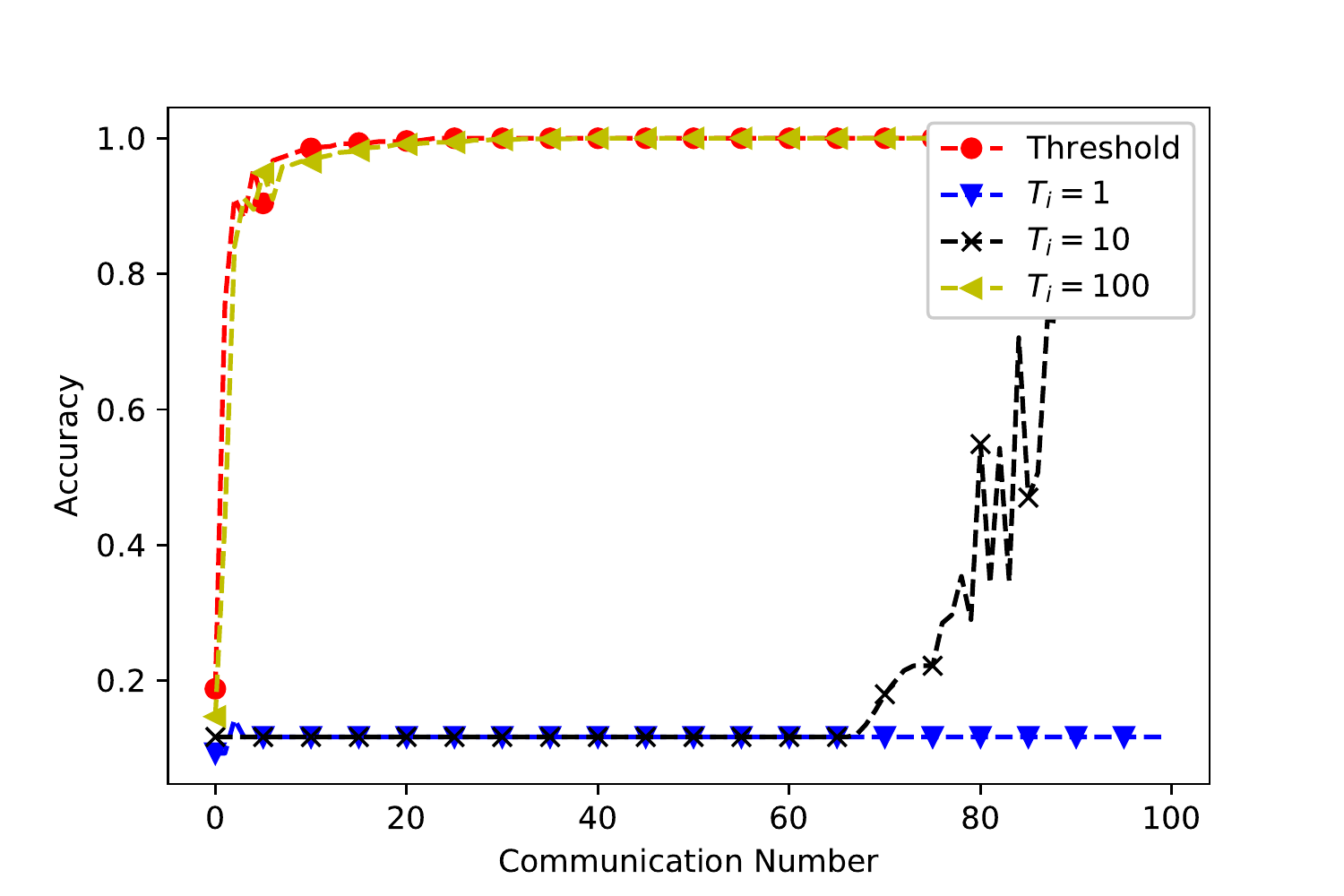}
	}
	\caption{LeNet on MNIST dataset with various node numbers.}
	\label{fig:LeNetMore}
\end{figure*}

We select the first 1000 instances from MNIST dataset and train LeNet with various node numbers $m=2,5$. Results in Fig ~\ref{fig:LeNetMore} are consistent with with $m=10$, namely a sufficiently large $T_i$ is required to obtain good training performance.

\begin{figure*}[!h]
	\centering
	\subfigure[Loss comparison]
	{\label{}
		\includegraphics[width=.42\linewidth]{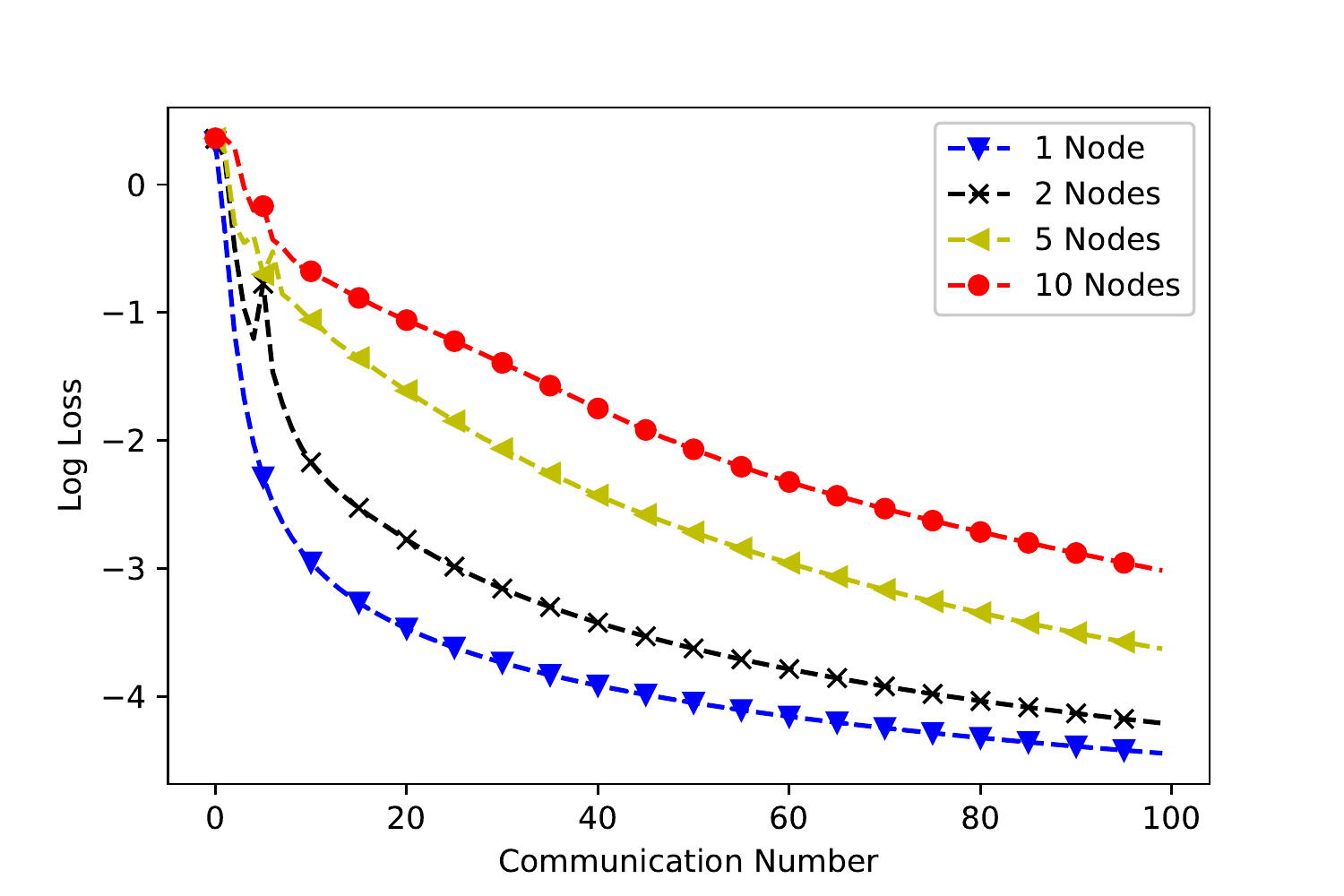}
	}
	\subfigure[Accuracy comparison]
	{\label{}
		\includegraphics[width=.42\linewidth]{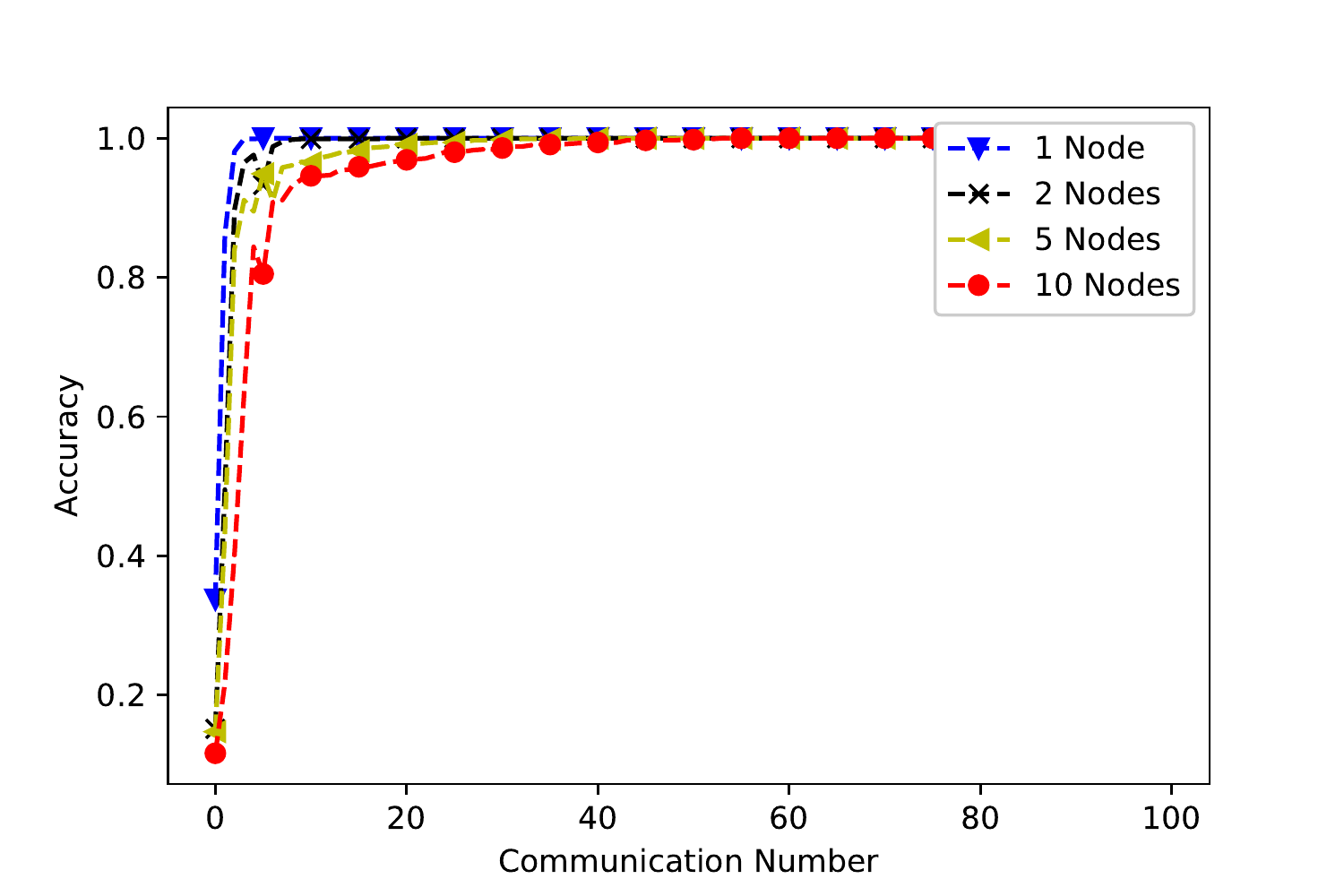}
	}
	\caption{LeNet on MNIST dataset with various node numbers. $T_i=100$ for all cases. }
	\label{fig:LeNetNode}
\end{figure*}

We then study how the node number $m$ affects the convergence rate by fixing $T_i =100$. Results in Fig ~\ref{fig:LeNetNode}  indicate that  more node numbers decrease the convergence rate.

\clearpage
\subsection{LeNet Experiments with Generalization Performance}
\begin{figure*}[!h]
	\centering
	\subfigure[Training loss with various $T_i$, $m=5$.]
	{\label{}
		\includegraphics[width=.42\linewidth]{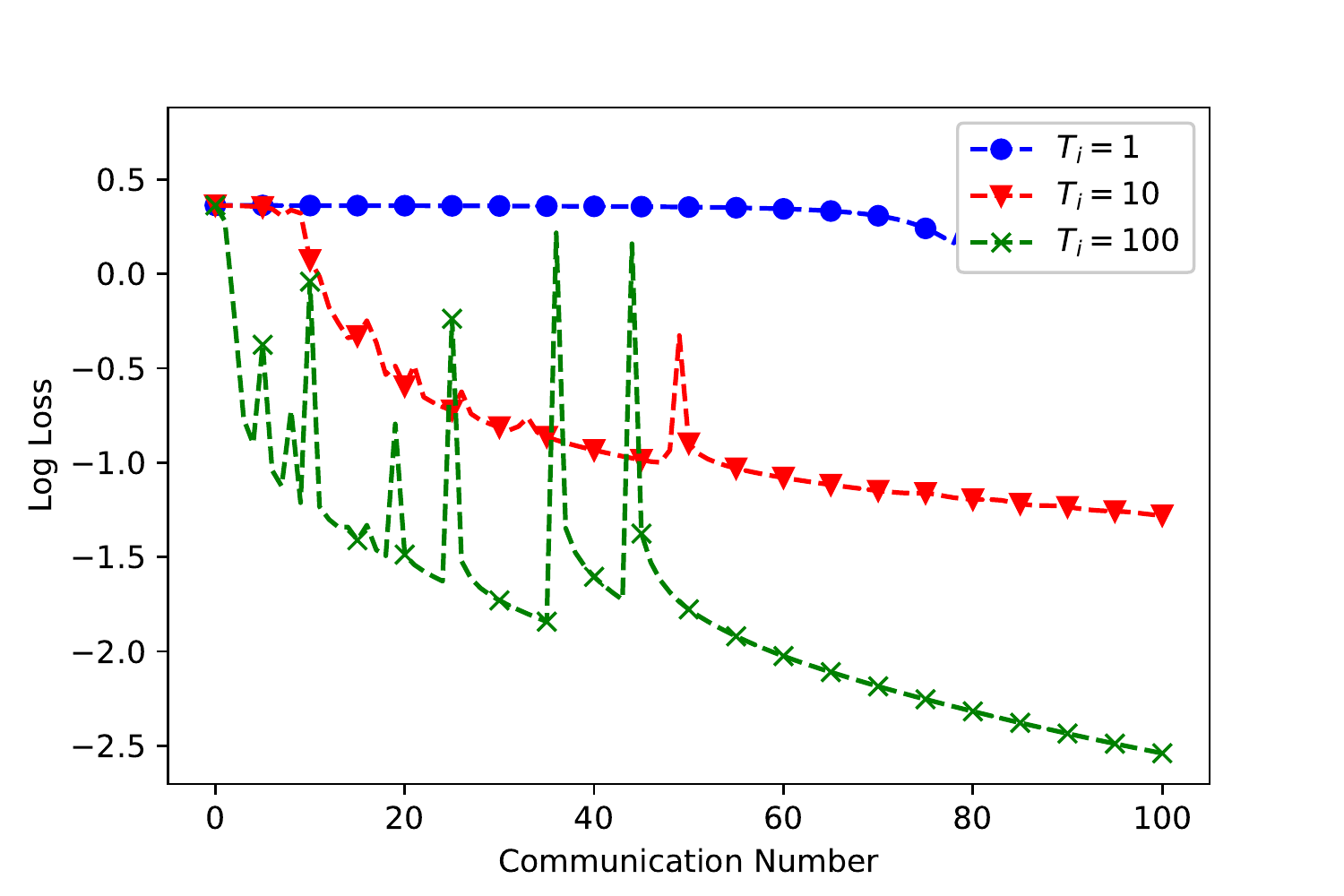}
	}
	\subfigure[Test loss with various $T_i$, $m=5$.]
	{\label{}
		\includegraphics[width=.42\linewidth]{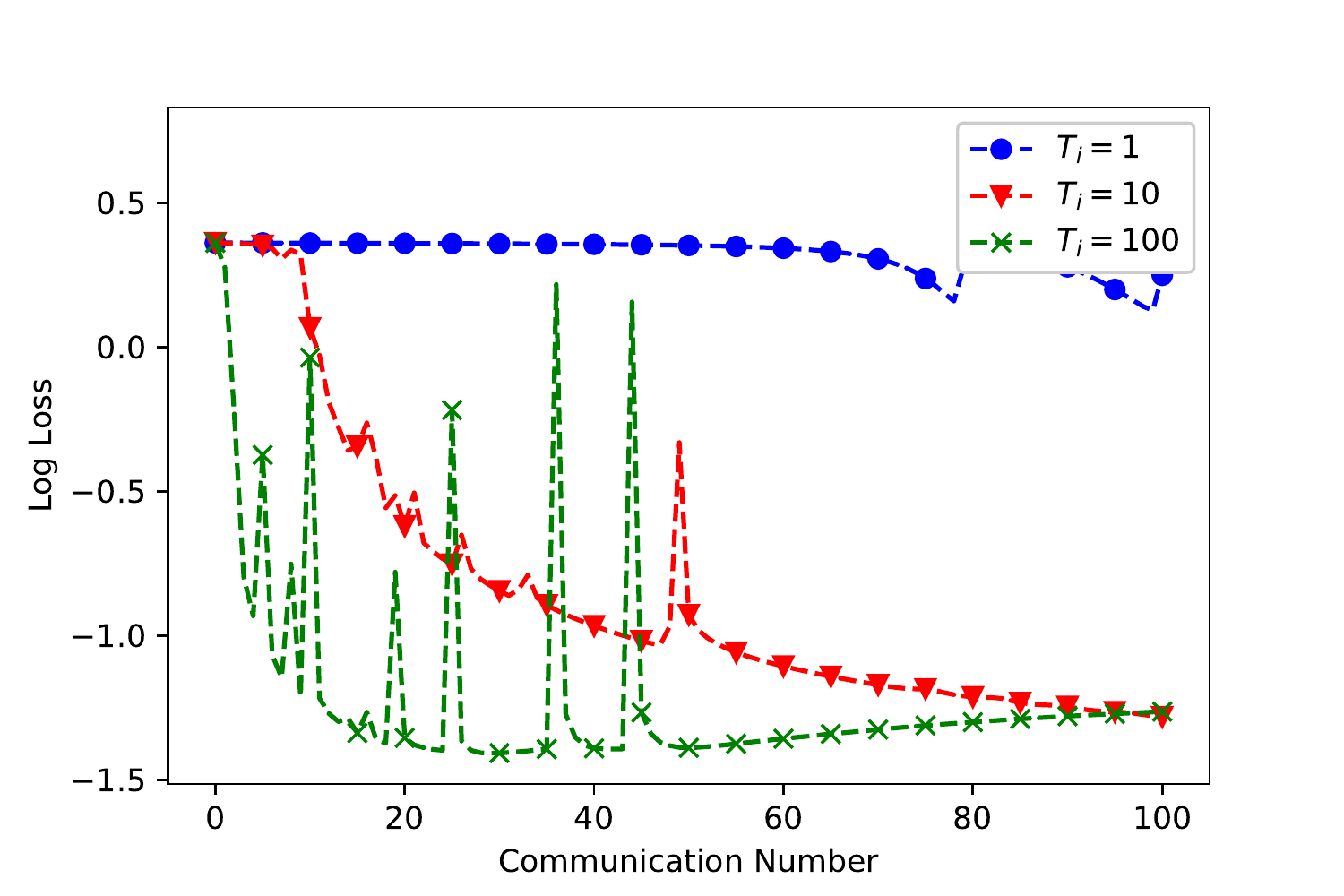}
	}
	\subfigure[Training loss with various $T_i$, $m=10$.]
	{\label{}
		\includegraphics[width=.42\linewidth]{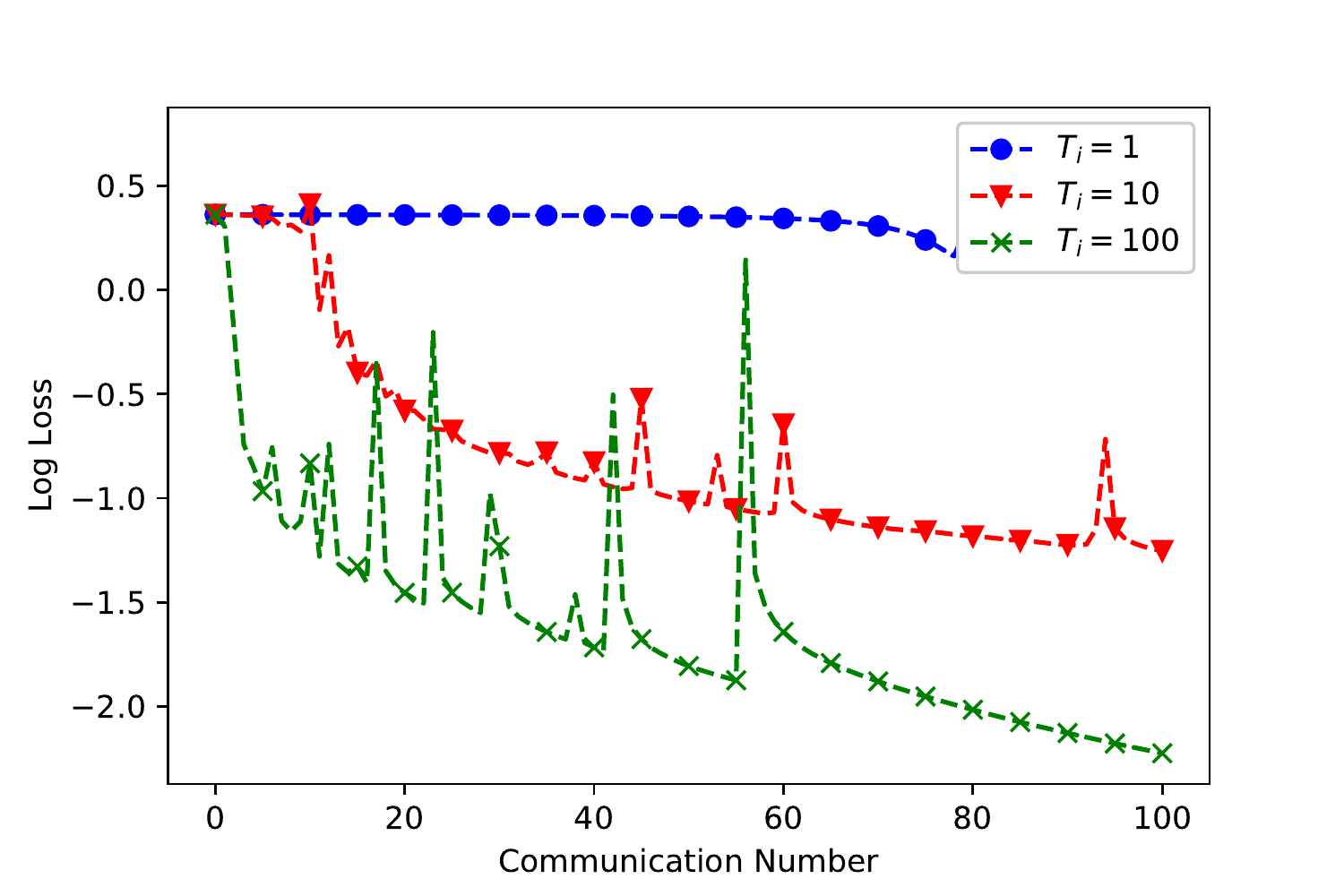}
	}
	\subfigure[Test loss with various $T_i$, $m=10$.]
	{\label{}
		\includegraphics[width=.42\linewidth]{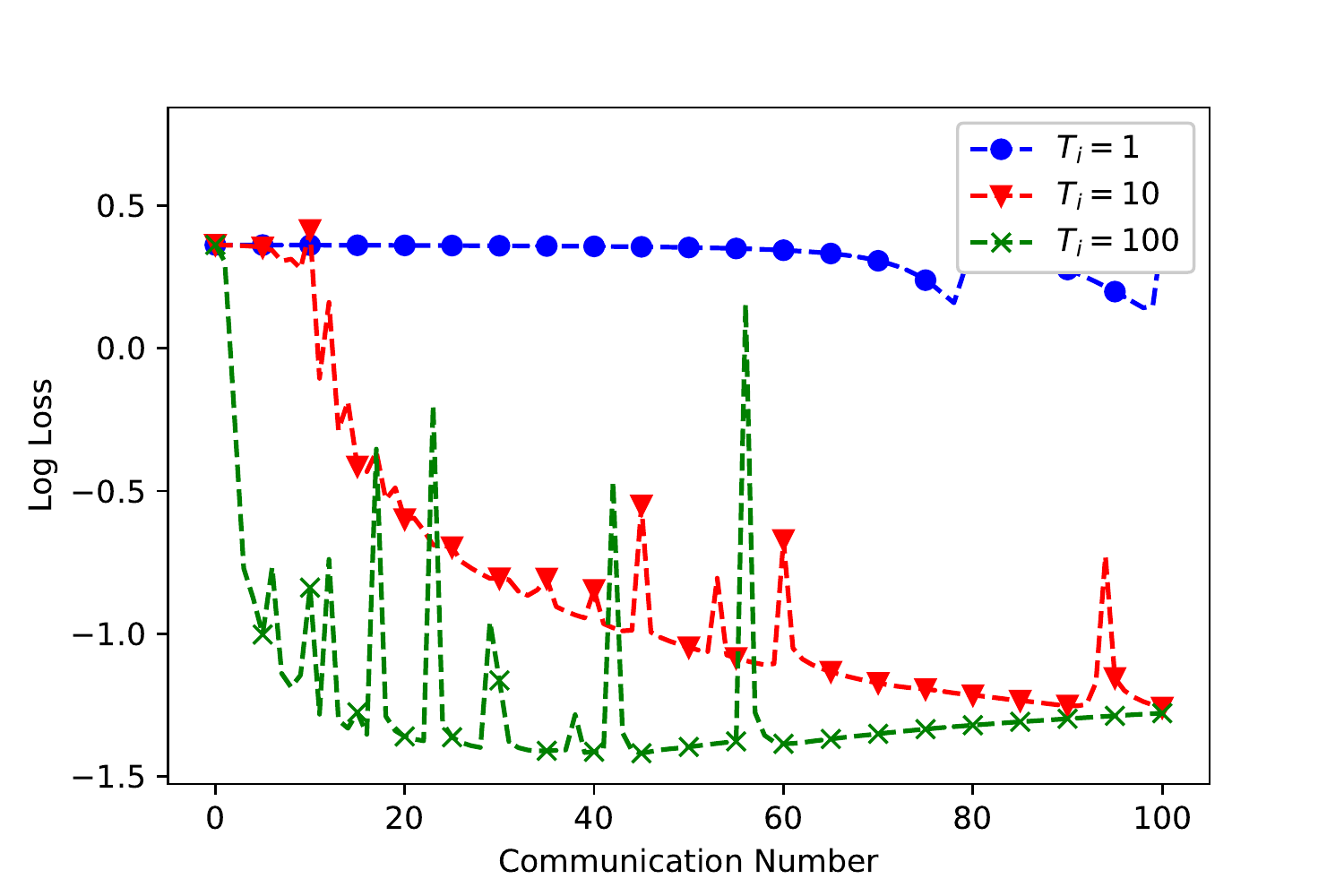}
	}
	\caption{LeNet on full MNIST dataset.}
	\label{fig:LeNetFull}
\end{figure*}

For completeness of our results, we also conduct experiments on the full dataset of MNIST. Results in Fig~\ref{fig:LeNetFull} are consistent with previous results, namely larger local updating $T_i$ decreases the training loss faster. Similarly, generalization performance is also better with larger $T_i$.

\end{document}